\documentclass[superscriptaddress, amsmath,amssymb, pre, twocolumn,floatfix]{revtex4-1}
\usepackage{graphicx}
\usepackage{dcolumn}
\usepackage{bm}
\usepackage{hyperref}
\usepackage{marginnote}
\usepackage{xcolor}
\usepackage{float}

\begin{document}
\title{Loops versus lines and the compression stiffening of cells}

\author{M. C. Gandikota}
\affiliation{Physics Department, Syracuse University, Syracuse, NY 13244 USA}
\author{Katarzyna Pogoda}
\affiliation{Institute for Medicine and Engineering, The University of
Pennsylvania, Philadelphia, PA 19104, USA}
\affiliation{Institute of Nuclear Physics, Polish Academy of Sciences PL-31342, Krakow, Poland}
\author{Anne van Oosten}
\affiliation{Institute for Medicine and Engineering, The University of
Pennsylvania, Philadelphia, PA 19104, USA}
\affiliation{Leiden Academic Centre for Drug Research, Leiden University, Leiden, the Netherlands}
\author{T. A. Engstrom}
\affiliation{Physics Department, Syracuse University, Syracuse, NY 13244 USA}
\author{A. E. Patteson}
\affiliation{Physics Department, Syracuse University, Syracuse, NY 13244 USA}
\author{P. A. Janmey}
\affiliation{Institute for Medicine and Engineering, The University of
Pennsylvania, Philadelphia, PA 19104, USA}
\affiliation{Departments of Physiology and Physics \& Astronomy, The
University of Pennsylvania,  Philadelphia, PA 19104, USA}
\author{J. M. Schwarz}
\affiliation{Physics Department, Syracuse University, Syracuse, NY 13244 USA}
\affiliation{Indian Creek Farm, Ithaca, NY 14850, USA}
\date{\today}

\begin{abstract}
Both animal and plant tissue exhibit a nonlinear rheological phenomenon known as compression stiffening, or an increase in moduli with increasing uniaxial compressive strain. Does such a phenomenon exist in single cells, which are the building blocks of tissues?  One expects an individual cell to compression soften since the semiflexible biopolymer-based cytoskeletal network maintains the mechanical integrity of the cell and {\it in vitro} semiflexible biopolymer networks typically compression soften. To the contrary, we find that mouse embryonic fibroblasts (mEFs) compression stiffen under uniaxial compression via atomic force microscopy studies. To understand this finding, we uncover several potential mechanisms for compression stiffening. First, we study a single semiflexible polymer loop modeling the actomyosin cortex enclosing a viscous medium modeled as an incompressible fluid. Second, we study a two-dimensional semiflexible polymer/fiber network interspersed with area-conserving loops, which are a proxy for vesicles and fluid-based organelles. Third, we study two-dimensional fiber networks with angular-constraining crosslinks, i.e. semiflexible loops on the mesh scale. In the latter two cases, the loops act as geometric constraints on the fiber network to help stiffen it via increased angular interactions. We find that the single semiflexible polymer loop model agrees well with the experimental cell compression stiffening finding until approximately 35\% compressive strain after which bulk fiber network effects may contribute. We also find for the fiber network with area-conserving loops model that the stress-strain curves are sensitive to the packing fraction and size distribution of the area-conserving loops, thereby creating a mechanical fingerprint across different cell types. Finally, we make comparisons between this model and experiments on fibrin networks interlaced with beads as well as discuss implications for single cell compression stiffening at the tissue scale. 
\end{abstract}

\maketitle

\section{Introduction}

Compression stiffening, a nonlinear rheological property in which a material's moduli increase with increasing uniaxial compressive strain, has recently been discovered in several types of animal and plant tissues~\cite{pogoda,perepelyuk,Engstrom19}. Some of these tissues contain a filamentous extracellular matrix, while others do not. Given these studies, a natural question emerges: Since individual cells are the building block of tissues, do individual cells compression stiffen? Should the answer to this question be affirmative, one cannot necessarily conclude that tissues compression stiffen given the possibility of emergent, collective mechanical phenomena, however, answering the question is surely a reasonable starting point.  Interestingly, we will explore the possibility of emergent mechanical phenomena within an individual cell. 

From a mechanical perspective, the cytoskeleton gives the cell its structural integrity. The cytoskeleton consists of actin filaments, intermediate filaments, and microtubules~\cite{alberts}, all of which are semiflexible biopolymers~\cite{gardel}. Semiflexible polymers have a characteristic persistence length $l_p$ such that for lengthscales much lower than $l_p$, they act as rigid rods, while for length scales much larger than $l_p$, they act as flexible (Gaussian) polymers. A typical persistence length for intermediate filaments is approximately 1 micron~\cite{mucke}, while for actin it is approximately 17 microns \cite{gittes1993,ott1993}. These semiflexible polymers crosslink to form a composite semiflexible polymer network. Actin dominates near the periphery of the cell \cite{alberts}. In contrast, vimentin, an intermediate filament, is localized more around the nucleus and other organelles to presumably anchor them in place~\cite{shabbir2014geometric,guo2013role}. Vimentin also enhances the elasticity of a cell with the enhancement increasing with increasing substrate stiffness~\cite{Melissa} as well as suppresses nuclear damage in cells undergoing large deformations~\cite{Patteson1}. 

If the mechanics of the cell is dominated by the cytoskeleton, then one can directly probe the mechanics of  {\it in vitro} semiflexible biopolymer networks to understand the mechanics of a cell. Such networks strain-stiffen under shear~\cite{janmey,storm}.  On the other hand, semiflexible biopolymer networks typically soften under compression~\cite{vanoosten}. Both mechanical responses are related to the mechanics of a single semiflexible polymer. An individual semiflexible polymer extension stiffens, i.e. its elastic modulus increases with extension strain, and compression softens, i.e. its elastic modulus decreases with compressive strain \cite{mackintosh1995elasticity, broedersz}. Stiff and semiflexible polymers compression soften as a consequence of the Euler-buckling instability with the transition being more gradual in the latter case due to the presence of fluctuations~\cite{landau,pilyugina}. Shear strain stiffening of semiflexible polymer networks is due to stretching out the polymers, combined with semiflexible polymers buckling orthogonal to the ones that stretch the most~\cite{janmey}.  In such systems, the filament density must be small enough to allow for the lengthening of the polymers. Compression softening at the network scale is attributed to filaments buckling, which then no longer contribute to the stiffness of the network as it is compressed~\cite{vanoosten}. 

If the cytoskeleton compression softens, such as {\it in vitro} semiflexible polymer networks do~\cite{vanoosten}, then how do cells protect themselves against compressive strains? Of course, cells are not just bags containing semiflexible biopolymer networks that can rearrange, they are also filled with vesicles and organelles.  Does the presence of vesicles and organelles then help protect the cell against compressive strains?  More specifically, if vesicles and organelles are modeled as regions of incompressible fluid, does the presence of such structures promote compression softening? Or, do they contribute to compression stiffening? And what about organelles that are elastic in nature? A majority of our modeling will focus on fluid-like organelles.  In addition, one can ask how does the typical mechanics of semiflexible biopolymer networks change in the presence of angle-constraining cross-linkers?  To date, most modeling has focused on freely-rotating crosslinkers~\cite{broedersz} with the exception of Refs.~\cite{das,lin,hatami}. With angle-constraining crosslinkers, one introduces semiflexible polymer loops at the network mesh scale. Unlike a semiflexible filament, a semiflexible loop does not buckle in plane and so one may expect the mechanics to differ. 

We will answer some of these questions by first conducting an experiment to determine whether or not cells compression stiffen or compression soften.  We will find that cells do compression stiffen, intriguingly. We will, therefore, investigate the role of vesicles and organelles embedded in a semiflexible polymer network (hereafter termed a fiber/fibrous network) and semiflexible polymer loops at the network mesh scale and at the cortex scale---to look for various mechanisms of compressional stiffening.  We will also study experimentally an {\it in vitro} fiber network embedded with beads so that we, in part, can more directly test ideas developed in our modeling. 

The paper is organized as follows.  We first present our experimental results, then we present our modeling and discuss how the modeling results can used to interpret the experimental results.  We conclude with a summary and discussion of implications of compression stiffening at the cell scale and how it may inform how compression stiffening occurs at the tissue scale~\cite{pogoda,perepelyuk}.

\section{Experiments}

To study the nature of compression stiffening in cells, we conduct two different experiments.  The first is whole cell compression of mouse embryonic fibroblasts (mEFs) and the second is compression of a fibrin network embedded with beads.  Since the cell contains both a boundary actomyosin cortex and a bulk fiber network, with the second experiment we are able to identify compression stiffening coming solely from a bulk fibrous network.

{\it Whole cell compression:} Experiments with whole cell compression of mouse embryonic fibroblasts (mEFs) were performed using a JPK Nanowizard 4 atomic force microscope equipped with cantilevers of a nominal stiffness of 2.4 N/m with a 25 $\mu$m diameter sphere attached (Novascan), according to a previously published protocol~\cite{hannah}  with minor modifications. Briefly, cells were trypsinized in order to round up and detach from the surface of the TC flask. Next, cells were centrifuged and resuspended in growing medium. Immediately round cells were placed on a Petri dish which was mounted on the AFM stage and indented uniaxially with a constant force of 450 nN at a speed of 5 $\mu$m/s as follows: (i) the AFM cantilever was placed over the rounded cell as controlled visually through the optical microscope, (ii) the point of contact between the cantilever and cell surface was recorded and assumed to be the cell height, (iii) each cell was indented until 450 nN force was reached and data were saved automatically as force (nN) vs. distance curves ($\mu$m). Such curves were then converted into stress (kPa) vs cell height (\%) with the assumption that normal stress can be calculated as the ratio of the applied force (F) to the area of deformation. The area of deformation $A$ was calculated as a spherical cap of the sphere, or $A=2\pi r h$ where $r$ is the radius of the sphere and $h$ is the depth at which cell was indented. The cell height percentage was calculated as the percentage of the total cell height that underwent indentation at a given force.  Assuming that the strain is 0\% at 100\% cell height, then the cell height can be converted to a strain percentage by subtracting the cell height percentage from 100\%.  Finally, the stress is then given by the ratio of the force to the area of deformation. The data was obtained from 10 cells and averaged over with the error bar denoting the standard deviation.  

\begin{figure}[h]
\includegraphics[width=0.48\textwidth]{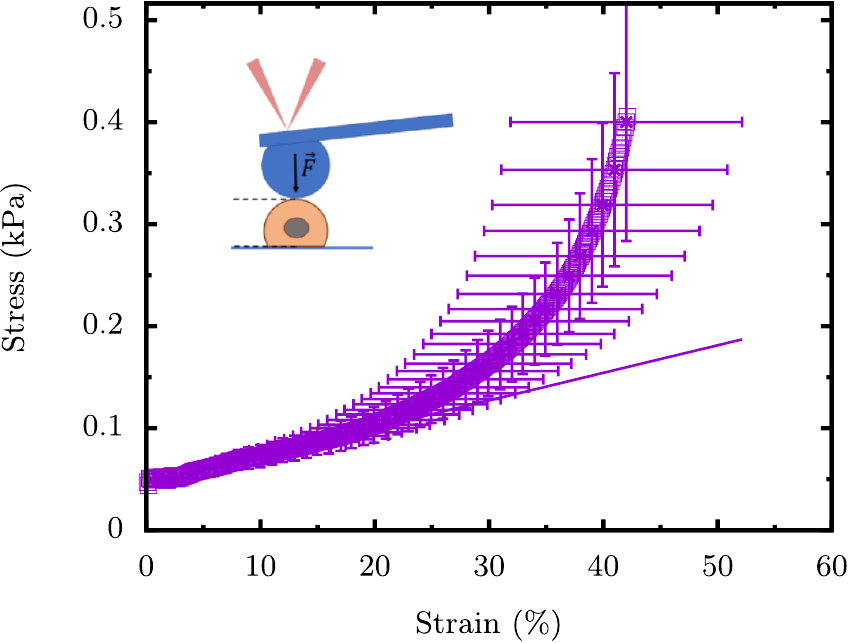}
\caption{{\it Compressive stress versus compressive strain for wild-type mouse embryo fibroblast cells.} The symbols represent the data and the line represents a linear fit to the data for up to 20\% strain. We observe the onset of compression stiffening around $\gamma_c\approx 20$\%. The inset is a schematic of the experiment where the AFM tip is attached to a glass bead (blue) which in turn applies a global strain on the mEF cell (salmon). The data is averaged over ten mEF cells with the error bars denoting the standard deviation.}\label{experiment}
\end{figure}

As evidenced by the stress-strain curve, these cells exhibit {\it compression stiffening} (see Fig. \ref{experiment}). Compression stiffening can be defined as a non-linear phenomenon in which the elastic modulus of the system increases with increasing compression, which is to be contrasted with uniaxially straining a Hookean spring where the spring constant remains independent of the strain. We define the strain at which the compression stiffening sets in as $\gamma_c$. See Table \ref{table_symbols} for the definition of this parameter and others used in the manuscript. The compression stiffening results of the mEF cell are a surprising mechanical response of the cell. The cytoskeleton, being a semiflexible polymer network is expected to compression soften due to the buckling of individual polymers. This disagreement between experiment and cell modeling necessitates a need to find new mechanisms for the observed behavior.  

{\it Fibrin network compression:}  The experimental protocol follows Ref.~\cite{vanoosten2} in which further experimental details are provided.

To study fibrin networks with embedded inert beads, fibrinogen isolated from human plasma (CalBioChem, EMD Millipore, Billerica, MA, USA) was dissolved in buffer. To prepare fibrin networks, fibrinogen, thrombin, 1X T7 buffer, and CaCl$_2$ solution were combined to yield 10 mg/mL fibrinogen, 30 mM Ca$^{2+}$, and 2 U/mL thrombin and allowed to polymerize between the rheometer plates for 1.5-2 hours at 37$^{\circ}$C and then surrounded with T7 buffer.  Beads made from cross-linked dextran (Sephadex G-25 medium, GE Health Sciences, Marlborough, MA) were swollen with H$_2$O to accomplish a 92\% swelling. The volume needed for 92\% swelling was extrapolated from the amount of water needed for 100\% swelling. The 100\% swelling was determined by allowing pre-weighed beads to swell for 12 hours in excess amounts of ddH$_2$O. The suspension was centrifuged at 2200 x g for 30 minutes, and the weight of volume of beads and excess water were determined.  

Fibrin networks with adherent beads: Fibrinogen 1 and thrombin 2 purified from salmon plasma (Sea Run Holdings, Freeport, ME, USA) were dissolved in 50mM Tris, 150 mM NaCl, pH 7.4 (T7 buffer).  Anion exchange chromatography beads (SP Sephadex C-25, GE Health Sciences, Marlborough, MA) to which fibrin binds were swollen to their equilibrium size in the same buffer.  

For rheometry, fibrinogen, T7 buffer, CaCl$_2$ solution, thrombin and water were mixed together first and then added to a bead solution to yield a fibrin network of the required concentration in a 1X T7 buffer with 0.5U thrombin/mL sample and the required volume of beads. Samples were polymerized between the rheometer plates for 90 minutes at 25$^{\circ}$C and surrounded by T7 buffer.   

\begin{figure}[h]
\includegraphics[width=0.48\textwidth]{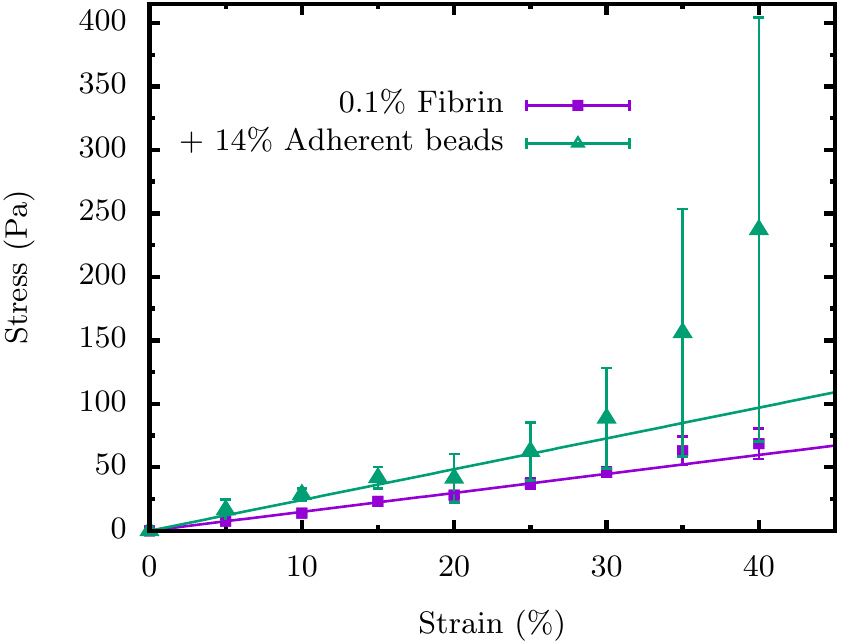}
\caption{{\it Compressive stress versus compressive strain for a fibrin fiber network with and without adherent dextran beads.} The symbols represent the data and the lines represent a linear fit to the data for up to 20\% strain. In the absence of beads, we do not observe compression stiffening. In the presence of 14\% packing percentage of adherent beads, we do observe compressional stiffening around $\gamma_c \approx 30$\%. Error bars denote standard deviation.}\label{experiment2}
\end{figure}

The experimental findings are as follows. Without beads, a 0.1\% fibrin network does not compression stiffen. However, even with just 14\% packing percentage of adherent beads, the fibrin network compression stiffens around 30\% compressive strain. See Fig. \ref{experiment2}.  This small packing fraction is far below both the packing percentage of random loose packing (55\%)~\cite{rlp} and random close packing (64\%)~\cite{rcp} of beads in three-dimensions. Thus the effect is not due to the jamming of the beads but rather an effect of the composite system.  With inert beads and a 1\% fibrin network, there is no compression stiffening until the packing percentage of beads is 60\% (See Ref.~\cite{vanoosten2}). 

\begin{figure*}[ht]
 \includegraphics[width=0.99\textwidth]{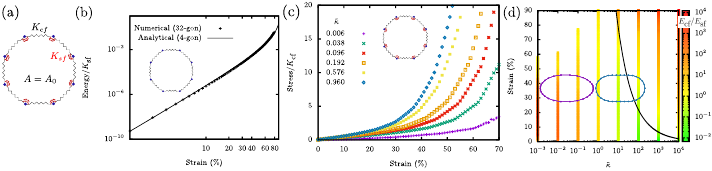}
\caption{{\it A cell as a viscous interior surrounded by an actomyosin cortex.} (a) The schematic of the system with central force spring (black) between neighboring vertices (blue) and angular spring (red) across a vertex. The spring constants are $K_{cf}$ and $K_{sf}$ respectively. The area of the polygon is preserved as the system is uniaxially compressed. (b) With just the central force springs, energy is seen to be quartic at small strain. Analytical calculations confirm the same (see Appendix (\ref{oneloop_appendix})). (c) Adding angular springs to the system brings linear behaviour at small strain since bending energy is quadratic at small strain (see Appendix \ref{oneloop_appendix}). This delays the onset of non-linearity effected by the central force springs. The onset of non-linearity is tuned by changing $\tilde{\kappa}$. (d) Heat map for the ratio of stretching and bending energy, $E_{cf}/E_{sf}$ as a function of $\tilde{\kappa}$ and strain. The solid black line is an analytical estimate separating the bending and stretching regimes. The shape of the polygon at 30\% strain for $\tilde{\kappa} = 0.006$ (dark-violet) is ellipse-like and for $\tilde{\kappa}=0.960$ (blue) is pill shaped. All numerical results were obtained using a 32-gon.} 
\label{cellasorganelles_fig}
\end{figure*}

\begin{table}[ht]
\begin{tabular}{l l l}
&Definition&Value\\
\hline
$\gamma_c$ &Strain at onset of compression stiffening& \\
$\sigma$&Compressive stress&\\
$K_{cf}$ &Central force spring constant& \\
$K_{sf}$ &Semiflexible angular spring constant& \\
$l_o$ & Distance between neighboring vertices&\\
&at zero strain &\\
$\tilde{\kappa}$ &Dimensionless constant - $\frac{K_{cf}l_0^2}{K_{sf}}$ &0.006 - 0.96\\
$K_{xlink}$ &Crosslinker angular spring constant&\\
$p$ &Bond occupation probability &0.5 - 1\\
$\phi$ &Packing fraction of area conserving loops &0.04 - 0.25\\
$\lambda$ &Lagrange multiplier &\\      
$K_A$ & Area-conserving loop ``spring'' constant &\\
$A_0$ & Preferred area &\\
\end{tabular}
\caption{Definitions of symbols.}\label{table_symbols}
\end{table}

\section{Cell as a viscous interior surrounded by a cortex}\label{One loop}
The simplest mechanical model for a cell is perhaps an actin cortex surrounding the periphery of the cell with an incompressible fluid inside. In other words, there is no rigid fiber network spanning across the cell and so we neglect its mechanical contribution. Without such a fiber network, organelles and vesicles remain disconnected at the cell scale and so act as viscous agents. For simplicity, we assume a two-dimensional geometry and will later address under what conditions is such a geometry applicable for a three-dimensional experiment. We model a cell as a loop (polygon) with a perimeter composed of springs that can stretch and bend and the polygon contains an incompressible fluid (see Fig. \ref{cellasorganelles_fig}a), i.e. the area enclosing the polygon is conserved. The Hamiltonian for a cell as a viscous interior surrounded by a cortex, $H_{v+c}$ with $v$ denoting viscous interior and $c$ denoting cortex, is then
\begin{align}\label{h0}
  H_{v+c} &= \frac{1}{2}K_{cf}\sum_{<ij>}(l_{ij}-l_o)^2 + \frac{1}{2}K_{sf}\sum_{<ijk>}(\theta_{ijk}-\theta_o)^2\nonumber\\
&+ \lambda\,(A-A_0), 
 \end{align}
where $l_{ij}$ is the length of a spring between vertices $i$ and $j$ and $l_0$ is the corresponding rest length. Additionally, $\theta_{ijk}$ is the angle between the polygon edges flanking the $j$th vertex and $\theta_o$ is its rest angle. Moreover, $A$ is the area of the polygon and $A_0$ is its preferred area, which is simply its initial area, and $\lambda$ denotes the Lagrange multiplier.  Finally, $K_{cf}$ and $K_{sf}$ denote the spring stiffness and bending stiffness respectively. 

At zero strain, a regular polygon of area $A_0$, is chosen as the initial configuration such that $H_{v+c}=0$, i.e. there is no pre-stress in the system. The vertices forming the polygon are then confined to be within two rigid lines. These lines are the two-dimensional equivalent of the compression plates in the experiment. Uniaxial compressive strain is applied by updating the position of the two parallel rigid lines and reducing the distance between them. Numerical minimization of the energy as defined in Eq.~\ref{h0} at various strains is performed using the SLSQP minimization algorithm in Python. This algorithm permits minimization while obeying strict constraints.  The compressive stress is defined as \begin{equation}\label{stress definition}
  \sigma = \frac{1}{A}\frac{\partial \bar{H}_{v+c}}{\partial \gamma}
 \end{equation}
where $\bar{H}_{v+c}$ is the numerically minimized energy at a given strain. 

For bending stiffness $K_{sf}=0$, by Maxwell constraint-counting of just the central-force springs, one would expect the loop not to be rigid at all for small strains~\cite{maxwell}. And yet, the energy of the polygon increases with  increasing strain (see Fig \ref{cellasorganelles_fig}b). This is solely due to the area conservation imposed on the loop during compression. Such a conservation can be thought of as exerting an outward ``pressure" onto the edges, making it untenable for the system to access its floppy modes. In the absence of bending, does such a loop compression stiffen?  We find a cubic stress-strain profile that can be understood via a minimal 4-polygon analytical calculation (see Fig. \ref{cellasorganelles_fig}b and Appendix~\ref{oneloop_appendix}) that makes an excellent fit to numerical results for higher polygons, i.e. 
\begin{equation}
\sigma \propto \gamma^3.
\end{equation}
In other words, the compressive strain at which the loop compression stiffens, $\gamma_c$, is zero in that the stress-strain curve is nonlinear for all strains. This cubic stress-strain curve is qualitatively different from the curves observed in Fig. ~\ref{experiment}.  This model, however, may be in agreement with the single cell compression experiments on T-lymphoma cells presented in Ref. \cite{lulevich}. In these experiments the cell is compressed between surfaces which are large compared to the dimension of the cell, thus the compression applies a global force on the cell. However, unlike compressive stress reported here,  Ref. \cite{lulevich} reports compressive force and fit their data using the above scaling. The authors show this fit to be good up to $\sim$ 30\% strain. 

While the $K_{sf}=0$ limit demonstrates compression stiffening, there is no linear stress-strain regime as observed for the MEF case. For $K_{sf}>0$, the perimeter of the polygon is now a stretchable semiflexible polymer. We do not consider buckling in our model since  semiflexible polymer loops with area conservation acting as a ``pressure", pushes the perimeter outwards, eliminating the possibility of buckling in this two-dimensional system.The competition between bending energy and area conservation has earlier been investigated in the context of finding the equilibrium shape of the loop \cite{arreaga}. Here, since an additional parameter $K_{sf}$ is introduced in the Hamiltonian, a tunable, dimensionless parameter $\tilde{\kappa}$ can now be defined.  Specifically, 
\begin{equation}\label{kappa tilda}
 \tilde{\kappa}=\frac{K_{cf}l_0^2}{K_{sf}}.
\end{equation}

Numerical minimization of $H_{v+c}$ shows compression stiffening with the added feature of having linear response at small strain (see Fig. \ref{cellasorganelles_fig}c).   The linear stress response at small strain is an outcome of adding angular springs to the polygon. An analytical calculation at small strains for this linear behavior for a 4-gon is presented in Appendix~\ref{oneloop_appendix}.  At larger strains, with growing compressive strain, the compressive stress increases more rapidly than a linear response. We see similar behaviour when a ``soft" area constraint is used in contrast to the ``hard" area constraint employed here. See Sec. \ref{oneloop_appendix}.

The energetics and the shape of the loop is determined by the dimensionless parameter $\tilde{\kappa}$ and the compressive strain $\gamma$. The heat map in Fig. \ref{cellasorganelles_fig}d studies the ratio of stretching to bending energy, $E_{cf}/E_{sf}$ as a function of both parameters. The black crossover line is obtained by equating the stretching and bending energies up to fourth order in the strain (see Eq. (\ref{diamond-stretch},  \ref{ang_expansion})). For $\tilde{\kappa}<1$, the system assumes an ellipse-like shape where angles are more conserved than distances between the vertices. Appendix \ref{oneloop_appendix} details a small strain calculation in the ellipse-like limit. For higher $\tilde{\kappa}$, i.e. $\tilde{\kappa}\approx 1$ and $\tilde{\kappa}>>1$, the system assumes a minimal pill shape in which distances between the vertices are more conserved. Pill-shaped surface have earlier been studied in the context of sea urchin eggs \cite{hiramoto1963}.

At low and medium strains in the heat map, $\tilde{\kappa}$ determines the domination of stretching or bending energy. For $\tilde{\kappa}<1$, the ellipse-like loop response is stretching dominated. For higher $\tilde{\kappa}$, the strain at which the pill-shaped loop transitions from bending to stretching is inversely proportional to $\tilde{\kappa}$. A larger $\tilde{\kappa}$ makes the loop less costly to bend and so bending energy contributes little to the total elastic energy. At a strain of around 40\%, the system's response to increasing $\tilde{\kappa}$ is stretching $\rightarrow$ bending $\rightarrow$ stretching dominated. This is distinctly different from shearing a fiber network where the system's response to increasing $\tilde{\kappa}$ is stretching $\rightarrow$ bending dominated~\cite{feng2016nonlinear}. Of course, the loop has a very simple network topology.  

At high strains, the system is stretching dominated for all $\tilde{\kappa}$. For $\tilde{\kappa}<1$, this is in line with the expectation for the ellipse-like loop. For higher $\tilde{\kappa}$'s, the angular springs of the pill-shaped loop that are in contact with the compression walls no longer contribute to the change in bending energy. The change in bending energy of the system then is proportional only to the number of vertices on the sides of the loop.  As the number of vertices on the sides of the loop decreases with strain, the elasticity of the system becomes increasingly governed by the stretching energy. Incidentally, the shearing of floppy fiber networks at large strains to induce rigidity appears to be stretching-dominated as well.

\section{Cell as a collection of organelles within a fiber network}\label{n loop}
We now ask how does the presence of a spanning, rigid fiber network affect the compression mechanics of a cell? While one cytoskeletal fiber type may not necessarily span the cell in a cross-linked network, a composite one is more likely to, particularly given the various means of couplings between the different filament types~\cite{fibercouplings}.  Since an individual {\it in vitro} fiber network typically compression softens, one anticipates that a composite fiber network compression softens as well, though we leave that as an open question.  For now, we look to other components of the cell to determine how they affect the mechanics.  Cells contain organelles that can be more fluid-based or more elastic in nature, and they contain vesicles. We will focus on the effect of fluid-based organelles and vesicles in this section and address elastic-based organelles in Sec. \ref{comp with expt}.   For simplicity, our modeling will be done in two-dimensions. Prior modeling has demonstrated that two-dimensional fiber network modeling qualitatively captures three-dimensional fiber network experiments~\cite{prestress}.  We will address the effect of dimensionality in Sec. \ref{comp with expt}. 

\begin{figure*}[ht]
\label{cellasnetwork}
\includegraphics[width=0.99\textwidth]{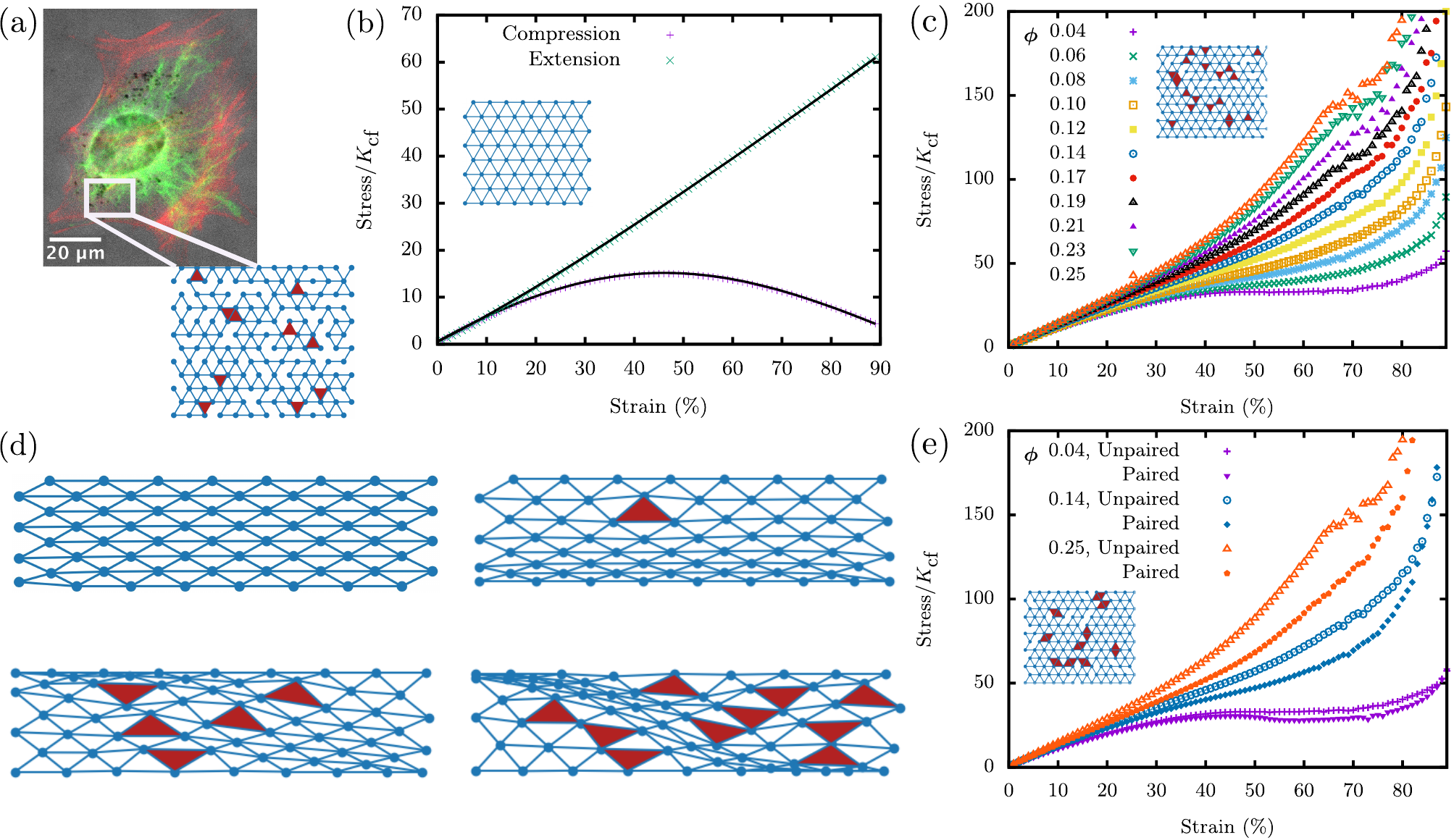}
\caption{{\it Cell as a collection of organelles within a fiber network.} (a) Immuno-fluorescence/phase contrast images of vimentin (green) and F-actin (red) in mouse embryonic fibroblasts adhered to glass slides demonstrating both a bulk fiber network and a boundary cortex. The dark spots are vesicles. The scale bar is 20 $\mu$m. We model the bulk fiber network as a randomly diluted triangular lattice. Each bond in the lattice denotes a central force spring. A pair of collinear bonds denotes an angular spring across its central vertex. Disorder is introduced by random dilution of bonds. Organelles are introduced via area-conserving loops, which are triangular in shape given the underlying structure of the lattice. (b) The fiber network compression softens in the absence of area-conserving loops. Data points are from simulations of a fully occupied lattice. Solid black lines are the analytical fits obtained by minimization of affine network energy (see Appendix \ref{network_appendix}) and are scaled here to fit with the numerical data.  The nonlinear response of the central force springs in the network causes compression softening, which is different and independent of the buckling mechanism. (d) For a given strain, the response of the network is influenced by the presence of area conserving loops. With no loops, an affine response is observed. Area-conserving loops influence the position and warping of the compressed layer. These non-affine deformations introduced by area conserving loops cause compression stiffening. (c, e) The size distribution of the area-conserving loops affects the elastic response of the system.  For a given packing fraction $\phi$, both networks have the same number of area-conserving loops, however (e) has the area-conserving loops linked together in pairs for three packing fractions. For comparison, three curves from (c) are also shown in (e). All numerical results were obtained using an occupation probability of $p=0.9$, $K_{cf}l_0^2/K_{sf}=1$ and curves were averaged over 100 runs on a 12x12 lattice.}\label{cellasnetwork_figure}
\end{figure*}

Therefore, we present a model with a network of fibers that are stretchable and bendable and with freely-rotating crosslinks. The fiber network also contains fluid-based organelles and vesicles as area-conserving loops randomly interspersed throughout. The compositeness of the cell focuses on the fibers and area-conserving loop mixture.  We work with a triangular lattice whose bonds can be diluted randomly and independently to become a disordered triangular lattice.  The fibers reside on the bonds of this lattice and the area-conserving loops are represented as triangles.  See Fig. \ref{cellasnetwork_figure}a. The Hamiltonian for the cell as a collection of organelles within a fiber network, $H_{o+fn}$ with $o$ denoting organelles and $fn$ denoting fiber network, is then     
\begin{align}\label{h2}
 H_{o+fn}&=\frac{1}{2}K_{cf}\sum_{<ij>} p_{ij}\;(l_{ij}-l_0)^2 \\
&+\,\frac{1}{2}K_{sf}\sum_{ijk=\pi}p_{ij}p_{jk}\;(\theta_{ijk}-\pi)^2 \nonumber\\
&+\, K_A\;\sum_{i'=1}q_{i'}(A_{i'}-A_0)^2.\nonumber
\end{align}
The two-body interactions of the central force springs with rest length $l_0$ is accounted by summing over neighboring indices $i,j$. The $p_{i,j}$s are random variables governing bond occupation and introduce disorder in the system. Specifically, $p_{i,j}$ is one with probability $p$ (or zero with probability $1-p$) signifying an occupied (or unoccupied) bond between vertices $i$ and $j$. The three-body interactions of the angular springs are accounted for by summing over three neighboring and collinear indices $ijk$. The rest angle of the angular spring is $\pi$, i.e. a straight fiber is the lowest bending energy configuration. The product of the random variables ensures that the bending term is non-zero only if both the central-force springs flanking the vertex are present. We work in the limit near $K_{cf}l_0^2/K_{sf}=1$ since bulk intermediate filaments, such as vimentin, are more stretchable than actin, for example~\cite{charrier}. Area-conserving loops are introduced as ``area springs" instead of using lagrange multipliers (as in Sec. (III)), the choice being made for computational simplicity. The area spring penalizes deviations from the preferred area $A_0$. To ensure that the area springs contribute only minimally to the total elastic energy, the area spring stiffness is set to be three orders of magnitude larger than the central force spring stiffness, i.e. $K_{A}l_0^2/K_{cf}=10^3$. We can then explore the effect of area-conserving loops on the mechanics of the fiber network. For each $i^\prime$th triangle in the network, $q_{i^\prime}$ is one with probability $\phi$ or zero otherwise. Here, $\phi$ is the packing fraction of the area conserving loops in the network. Finally, we implement open boundary conditions with the vertices constrained between two rigid lines. As before (see Sec. \ref{One loop}), the network is not prestressed initially and $H_{o+fn}$ is minimized for different compressive strains to obtain the stress-strain dependence.

\textit{With $K_A=0$}, we begin with no organelles and look for compressional softening.  We must emphasize that we have not implemented buckling at the single fiber level. Instead, we seek a more collective compression softening mechanism.  To do so, we begin with an ordered lattice ($p = 1$) where we see the network exhibit an affine response under compression and extension. In the affine regime, straight fibers in the network remain straight fibers and thus angular springs do not contribute to the elastic energy. We numerically find that the compression response of the network is in sharp contrast to extension, the latter of which remains linear throughout. See Fig. \ref{cellasorganelles_fig}b for the stress-strain curves. More specifically, the fiber network compression softens. 

A physical explanation for the softening is that when the network is compressed, the springs increasingly align along the transverse axis of compression. It is then easier to compress the system at larger strains for this given choice of orientation of the triangular lattice. See Appendix \ref{network_appendix} for the details of an analytical calculation quantifying the compression softening. Is this softening generic? For a triangular lattice rotated by 90 degrees, there would be no compressive softening since in the direction of compressive strain there would always be springs co-linear with the compression axis.  In contrast, for any rotation less than 90 degrees, there will be compression softening since there are no springs co-linear with the compression axis. Therefore, the 90 degree rotation is a singular case and not generic (see Fig. \ref{comp_soft_angle_dependence}). 

This compression softening phenomenon is \textit{independent} of the buckling of semiflexible polymers, which until now has been considered to be a dominant reason for compression softening of such networks. The signature of compression softening is also observed for disordered lattices with $p < 1$ (see Fig. \ref{comp softening} in Appendix \ref{network_appendix}).  We also note that this softening is distinct from the mechanism of mechanical collapse studied in central-force networks under biaxial compression in which a martensite-like transition occurs during the collapse~\cite{discher1997,wintz1997}. This martensite-like transition occurs in the absence of semiflexibility and in general when $K_{sf}<<K_{cf}l_0^2$. Compression softening has earlier been observed in a tensegrity model of a cell \cite{beysens2013dynamical}.

What can we expect when we include organelles as area-conserving loops into the fiber network? While working with our initial cell as a viscous medium surrounded by an actomyosin cortex, we saw that despite a loop of central-force springs being floppy according to Maxwell constraint counting, the area-conserving loop creates nonlinear rigidity as evidenced by the compressional stiffening with $\gamma_c=0$. The addition of bending leads to a linear regime at small strains. Will adding area-conserving loops to the fiber network do the same even if they are only few in numbers?  There are two competing factors at work here - the network's compression softening and the area-conserving loop's compression stiffening. We now investigate this competition by varying $\phi$, the packing fraction of area-conserving loops.

\textit{With $K_A>>0$}, area-conserving loops break the affine response of the network. A force balance argument (see Appendix (\ref{network_appendix})) shows that an area-conserving loop necessitates the angular springs around it to bend to ensure local mechanical equilibrium. Angular springs earlier did not contribute to the elastic energy in the affine response of the fiber network with no area-conserving loops. Given the non-affine deformations introduced by the area-conserving loops, angular springs begin to contribute to the total elastic energy of the system. To see bending modes in the network as the fibers bend to deform around the ``obstacle'', if you will, see Fig. (\ref{cellasnetwork_figure}d). These bending modes, therefore, lead to a compression \textit{stiffening} response (see Fig. \ref{cellasnetwork_figure}c) as the ``obstacles'' prevent the collapse of three springs along the three lattice lines of the triangular lattice onto one line.  The affine stretching-led compression softening competes with the non-affine bending-led compression stiffening.  This argument is independent of system size and so we have checked that this mechanics persists in both smaller and larger systems (see Fig. \ref{finite_size_inclusions} in Appendix (\ref{network_appendix})). 

If the cost of bending is too large, the area-conserving loops will simply deform even for small strains and the fiber network will remain affine even at large strains so that the bending contribution must not be much greater than the stretching contribution in order to observe this compression stiffening.  On the other hand, if the cost of bending is too small, then the fibers will easily deform around the organelles.  This energetic contribution may or may not be enough to combat the compression softening due to the stretching.  So the compression stiffening robustly occurs in the regime when bending energy is comparable to the stretching energy. 

Interestingly, even a few area-conserving loops ($\phi=0.04$) are sufficient for the angular springs to subdue the compression softening of the fiber network (see Fig. \ref{cellasnetwork_figure}c). With more area-conserving loops, the fibers are forced to bend more and therefore contribute to  compression stiffening of the fiber network at smaller strains.  It is additionally observed that the stress response is not just determined by the number of area-conserving loops in the network but also by their \textit{size distribution}. Keeping the packing fraction $\phi$ constant and now pairing up the area-conserving loops, the stresses are not as large in comparison to a network whose area-conserving loops are randomly distributed (see Fig. \ref{cellasnetwork_figure}e).  Since the stresses are not as large, the onset of the compressional stiffening is delayed to a larger $\gamma_c$.  This pairing up localizes the area-conserving loops as compared to the un-paired case such that there are effectively fewer obstacles to distort around.  Therefore, the stress-strain curves are not only sensitive to the packing fraction of the fluid-based organelles and vesicles but also to their size distribution.  In other words, the stress-strain curves are a mechanical fingerprint of the innards of a cell and one can study how the size distribution of such structures affects the mechanics.

Our small strain, affine, stretching deformations versus large strain, non-affine, bending deformations should be contrasted with earlier modeling of fiber networks. These earlier small strain studies demonstrate a change from affine, stretching dominated response to non-affine, bending dominated response as the fiber network is diluted~\cite{head2003deformation,broedersz2011criticality,das}. A similar change occurs by decreasing shear strain in substatic fiber networks, yet the strains at which the change occurs are large~\cite{sharma}. In this work, we observe that a stretching-to-bending change can be tuned by increasing the number of area-conserving loops. However, since we have reported results in superstatic networks, even in non-affine response, the energy is not dominated by bending, but bending only becomes comparable to stretching. 

\begin{figure}[h!]
 \includegraphics[width=0.48\textwidth]{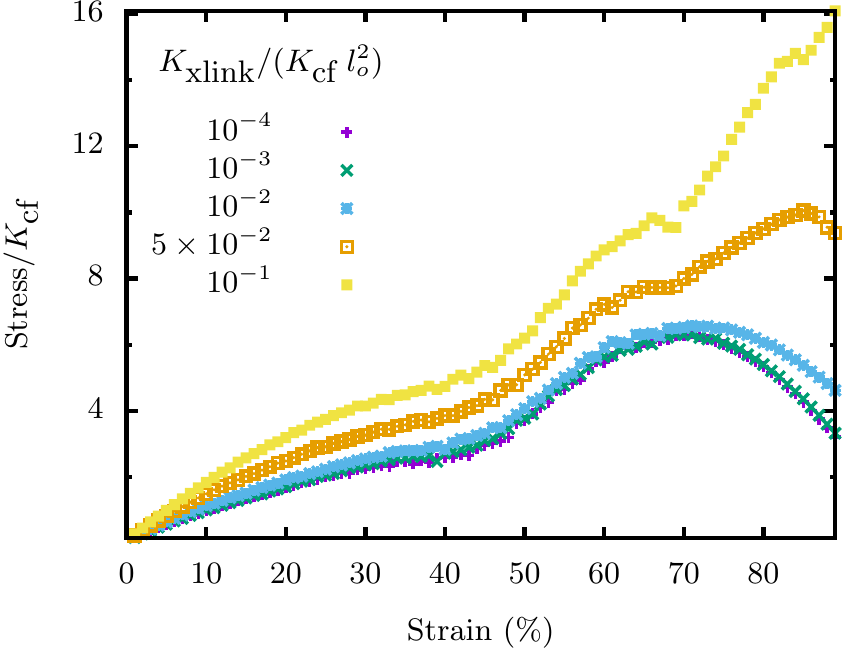}
 \newline
\includegraphics[width=0.48\textwidth]{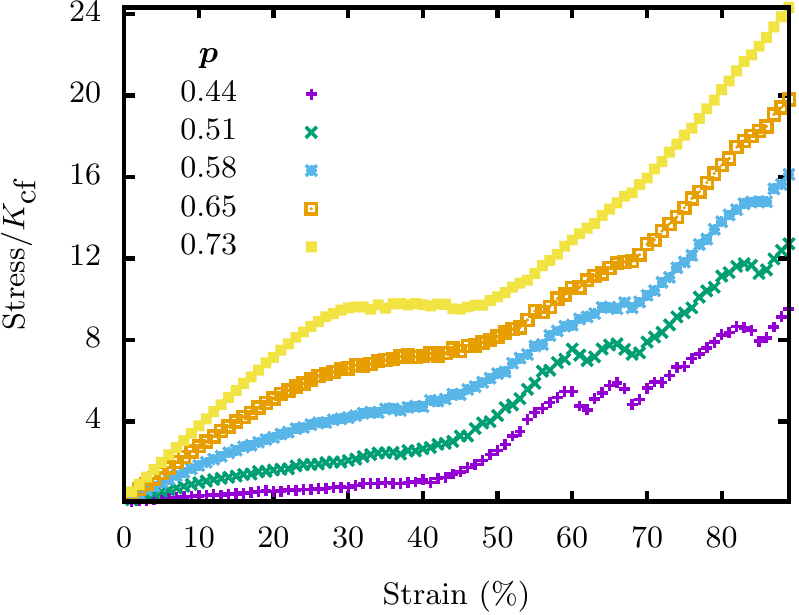}
 \caption{{\it Compression stiffening in networks with angle-constraining cross-links.} Area-conserving loops are absent in these networks. \textit{Top}: Compression stiffening as a function of $K_{xlink}/(K_{cf}l_0^2)$ is shown for systems with occupation probability $p=0.58$. When the ratio is small, $H_{fn+axlinks}$ reduces to $H_{o+fn}$ (with $K_A=0$) and compression softening is observed as expected. The onset of nonlinearity is not tunable by this ratio. \textit{Bottom}: With $K_{xlink}/(K_{cf}l_0^2)=0.1$, compression stiffening for different occupation probabilities $p$ is shown. For both figures, the curves are averaged over 1000 runs on an 8x8 lattice with $K_{cf}l_0^2/K_{sf}=1$. \label{kcross}}
\end{figure}

{\it Angle-constraining crosslinks:}  Before addressing the experimental data, let us briefly consider another potential mechanism for compression stiffening, namely angle-constraining crosslinkers.  Having such crosslinkers will, again, prevent the collapse of three springs into one spring perpendicular to the compression axis because the collapse is energetically unfavorable even in the absence of organelles. The Hamiltonian of such a fiber network with angle-constraining crosslinks is given by 
\begin{equation}
 H_{fn+axlinks}=H_{o+fn} + \frac{K_{xlink}}{2} \sum_{ijk=\frac{\pi}{3}}p_{ij}p_{jk}\;(\theta_{ijk}-\pi/3)^2,
\end{equation}
with $K_A=0$. Here, $K_{xlink}$ is the bending stiffness of the crosslinker spring and $\pi/3$ is the rest angle of the spring since we work on a triangular lattice. The response of such networks to shear strain has been studied~\cite{das,hatami,lin} but not in response to compression. This Hamiltonian corresponds to having non-area conserving semiflexible polymer loops at the mesh scale of the fiber network. 

In response to compressive strain, even without any area-conserving loops, this network compression stiffens as can be inferred from Fig. \ref{kcross}.  
At small strain, the angles between fibers change within each triangular loop with both stretching and angle-constraining crosslinks dominating the response.  At larger strains, the affine stretching eventually compression softens while the angle-constraining crosslinks become increasingly distorted to compression stiffen. When $K_{xlink}/K_{cf}l_0^2<<1$, the stretching-dominated compression softens wins.  However, as the ratio increases, eventually the bending-dominated compression stiffening wins.  Note that bending along fibers does not play much of a role here. See Fig. \ref{axlink_contributions}.  

We also explore the fiber network mechanics for different occupation probabilities with $K_{xlink}/(K_{cf}l^2_0)$ closer to unity.  Note that we can explore a larger range of occupation probabilities than with freely-rotating crosslinks because the $p$ above which the network is rigid is the connectivity percolation threshold for the triangular lattice, i.e. $p_c=2\sin(\frac{\pi}{18})=0.347$~\cite{das}.  For a range of $p<0.7$, the network response is similar.  However, for $p>0.7$, we observe a plateau in the stress-strain response occuring at intermediate strains. This plateau corresponds to a global distortion of the lattice to weaken it.  This phenomenon may be related to a first-order transition in the collapse of the network that was studied in Refs.~\cite{discher1997,wintz1997}, though with bending replacing stretching.  When $p=1$, there is a dramatic increase followed by a sudden decrease in the stress-strain relation at these intermediate strains.

\section{Comparison with experiments}\label{comp with expt}

Our experiments demonstrate that mEF cells exhibit compression stiffening with $\gamma_c\approx  20$\%. From the modeling side, we have identified three possible routes to compression stiffening in cells, namely, (i) a boundary actomyosin cortex enclosing a viscous medium in the absence of a bulk spanning fiber network, (ii) a bulk spanning fiber network with freely-rotating crosslinks and interspersed with fluid-based organelles and vesicles, and (iii) a bulk spanning fiber network with angle-constraining crosslinks. All three mechanisms produce a linear stress-strain relation at small strains before compression stiffening at strains larger than $\gamma_c$. For the bulk fiber network results, the compression stiffening finding is robust when bending is comparable to stretching. 

Which model is most relevant for the experiment at hand? If there is no bulk, rigid cytoskeletal network in mEFs, then one expects that the boundary cortex enclosing a viscous medium to be the most relevant model, at least up to strains where the nucleus is not in direct contact with the compression apparatus since the nucleus is typically the stiffest organelle in the cell~\cite{lammerding}.  This model is also consistent with studies of local and global cell stiffness using multiple methods that consistently show apparent Young's moduli of a few Pa in the cell interior but moduli in the range of kPa at the cell cortex\cite{wirtz}. If there is a bulk, rigid cytoskeletal network, then one expects organelles and vesicles embedded within a freely-rotating crosslinked fiber network to be the most relevant. Angle-constraining crosslinks can help amplify the effect.  Since we do not know directly whether or not there is a bulk, rigid cytoskeletal network, let us assume there is not and explore what our two-dimensional boundary cortex enclosing a viscous medium model can tell us about our three-dimensional experiment. 

Given the simpler two-dimensional modeling versus the three-dimensional experiment, one does not necessarily anticipate quantitative agreement between the two.  We now make a case for the potential for quantitative comparison.  Let us assume that the presence of the compression apparatus breaks the spherical-like symmetry of the cell and so it can be treated as a collection of two-dimensional cross-sections with minimal fluid flow between the cross-sections as the cell is compressed. Then the actomyosin cortex is captured by a loop and the volume conservation due to the viscous medium translates to area conservation within each cross-section. In addition, energy has the same units in any dimension, while stress does not. More precisely, the difference between a two-dimensional stress and a three-dimensional stress is simply a length factor. Alternatively, we can rescale the experimental results by a particular value to nondimensionalize the experimental results. 

Should the presence of the compression apparatus not break the spherical-like symmetry of the cell, if we consider the actomyosin cortex as a discrete set of loops that are connected together at various points to form a shell, then the compressional stiffening would then be dominated by the loop that is most likely to compression stiffen first. This argument, again, points to our loop model as a potentially accurate description of the mechanics, as long as the coupling between loops is weak. Should the coupling between loops be strong, then a full three-dimensional model consisting of multiple loops is needed. Within a multiple loops framework, fluid flow amongst the different loops (yet with overall volume conserved) results in a change in the area of the loops.  Since we do not yet know if cross-sections of the cell change in area as compression occurs, we cannot yet rule out the role of fluid flow within the cell.  Since we cannot rule out fluid flow, we can easily extend the two-dimensional loop model with conserved area to a loop model in which area is not-conserved by introducing a soft area constraint to account for the possibility of a cross-section of the cell changing area as it is compressed.  See the Appendix \ref{soft_area} for details. 

\begin{figure}[h!]
\includegraphics[width=0.48\textwidth]{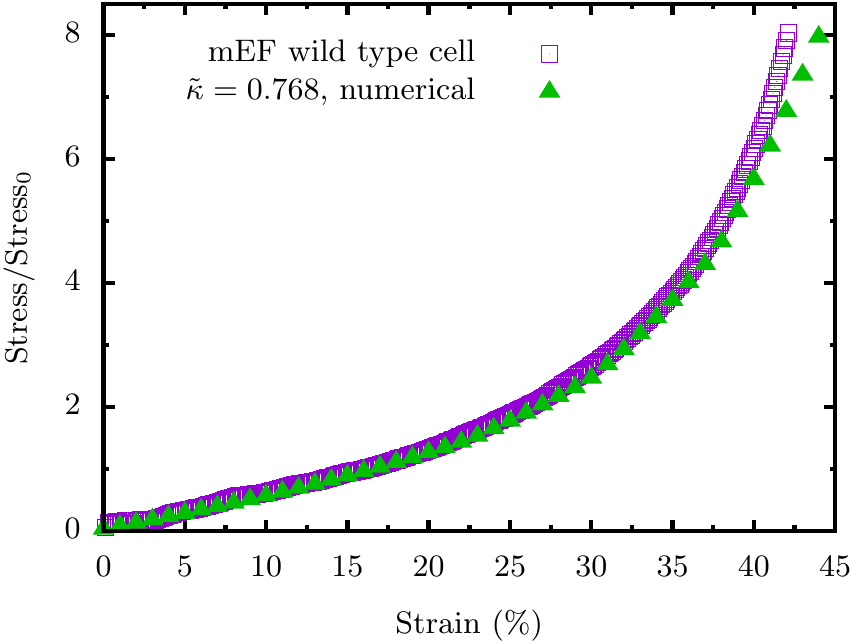}
 \caption{{\it Comparison of the experimental Figure 1 with a modeling curve from Figure 3c}. We subtract the pre-stress from the experimental curve and normalize the stress by the same value. Note that there is only one free parameter in the modeling curve, $\tilde{\kappa}$.}\label{oneloop_exp_comparison}
 \end{figure}

If we model the actomyosin cortex as a discrete set of multiple loops (spherical symmetry-breaking or not), there is the potential for quantitative comparison between our modeling and our experiments. We, therefore, present quantitative comparison between the experiment (from Fig. \ref{experiment}) and the modeling (from Fig. \ref{cellasorganelles_fig}c) in Figure \ref{oneloop_exp_comparison} in which the area is conserved. After subtracting the pre-stress from experimental data and using the same value to nondimensionalize the stress, we plot both curves on the same plot to obtain very reasonable agreement between the experiment and the model with only one free parameter, $\tilde{\kappa}$, that is somewhat constrained by earlier experiments. With $\tilde{\kappa}=0.768$, it is a regime in which both stretching and bending energy contribute to the total elastic energy of the cortex. Additionally, the ratio of bending energy to total elastic energy decreases monotonically with strain and the geometry of the cell is pill-shaped. The loop model with a soft area constraint also fits well with the experimental curves (see Appendix \ref{soft_area}), which means we cannot rule out either approach (area conserved or area not conserved) with the stress-strain curve alone. 

So while our boundary cortex enclosing a viscous medium appears to be a reasonable model, at least up until approximately 35\% compressive strain, the model begins to deviate from the experiment slightly beyond 35\% compressive strain. What other effects are at play?  Fibroblasts contain actin, vimentin, microtubules as well as other cytoskeletal fibers and they contain organelles such as the nucleus.  With large enough compression, the plates will encounter resistance from the stiff cell nucleus. For instance, for a nucleus that is one-quarter the volume of the cell, the strain at which the nucleus begins to dominate would be around 75\%. In addition, the more compressed a cell is, the more likely the fiber network will percolate across the cell at least in the direction perpendicular to the compression. Therefore, one cannot necessarily rule out the bulk fiber network interspersed with organelles model for larger strains. Moreover, given that we observe compression stiffening with angle-constraining crosslinkers even in the absence of organelles, the presence of angle-constraining crosslinkers, such as filamin A~\cite{filaminA}, only enhances such an effect. 

At this point, we also cannot necessarily rule out any of the compression stiffening mechanisms we have just presented. Perhaps all are at play at some level when compressing various cell types.  However, we can more directly compare our fluid-based organelles within a fiber network model if we compare to a reconstituted network of dextran beads embedded within an {\it in vitro} fibrous fibrin network with one modification to the model~\cite{vanoosten2}. Since the dextran beads essentially act as rigid objects even at 40\% strain, we modify the model accordingly by assigning the spring constant for the central-force springs surrounding any area-conserving loops to be 100 times larger than $K_{cf}$ to more closely approximate the rigidity of the beads. We have also added some small random variation in $K_{cf}$ for the central-force springs not surrounding a loop to more closely mimic the more generic network structure of the experiment that is presumably not based on an underlying lattice. 

\begin{figure}[h!]
\includegraphics[width=0.48\textwidth]{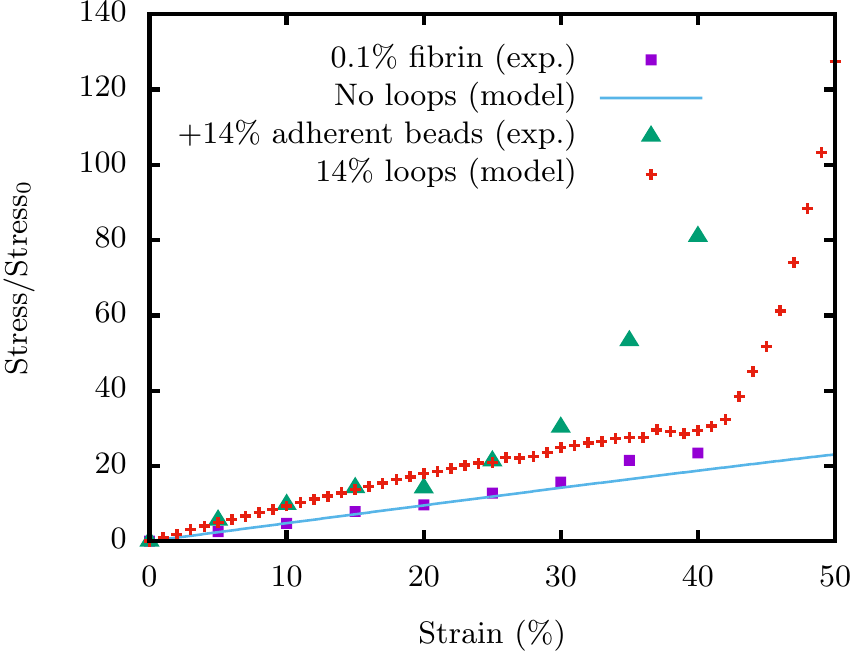}
 \caption{{\it Comparison of experimental Figure 2 with a version of Figure 4c.} The experimental curves are normalized axial stress versus axial compressive strain for the {\it in vitro} fibrin network with and without beads embedded. The modeling curves  are for  fiber networks with and without rigid area-conserving loops embedded. For the fiber network with no loops, the analytical curve for a central force spring with initial orientation of angle $\pi/2.2$ with respect to transverse axis of compression is chosen, see Fig. \ref{comp_soft_angle_dependence}.}\label{fibrin_fits}
\end{figure}

Is our fiber network with area-conserving loops model useful for interpreting these experimental results? Recall that the compression stiffening mechanism is robust when the stretching of fibers is comparable to the bending of the fibers.  Typically, individual fibers such as actin and collagen are not in this regime, though bundles of such fibers that can slide past each other may be closer to this regime. However, fibrin is a fiber with extraordinary extensibility and elasticity~\cite{fibrin,kim} making it a more likely candidate to be in such a regime. Figure \ref{fibrin_fits} is a dimensionless presentation of the data in Fig. \ref{experiment2} and demonstrates the comparison between the modeling and the experiment. Both experimental curves have been rescaled by 2.93 Pascals so that there would be one common data point between the modeling curves and the experimental curves at a strain of 10\%. As with the cell, if uniaxial compression of the initially isotropic system allows one to consider the composite system in terms of two-dimensional cross-sections, then our modeling is quantitatively applicable. Let us assume so given our cell results and discuss the comparison.  

Both the experiment and the model do not exhibit compression stiffening in the absence of beads/area-conserving loops. In the experiment with beads, compression stiffening occurs around 30\% strain and then by 40\% strain, the stress has increased about four-fold.  In the model with 14\% packing percentage of more rigid area-conserving loops, the onset of compression stiffening occurs around 42\% strain with a four-fold increase in stress by around 50\% strain.  This range can be modified by changing the spring constant stiffness of the central-force springs surrounding the area-conserving loop.  

So the most significant difference is the $\gamma_c$ in the experiment and the model. This difference may be due to a difference in length scales.  In the experiment, the bead diameter is much larger than the mesh size, while in the model, the two length scales are the same.  We expect more localized bending with smaller loops and less localized bending around bigger loops with both effects leading to compression stiffening, though how $\gamma_c$ is affected is not immediately clear. This expectation can be numerically tested by exploring larger loops embedded in larger lattices. There is an additional computational issue. With more rigid loops, they are more likely to overlap given their lack of deformability as the compression occurs, even in a nearly fully occupied triangular lattice. This overlap induces an unphysical softening in the model.  One can ameliorate this issue with vertex and edge annihilation and edge reassignment and would presumably shift the model's $\gamma_c$ to be more in line with the experiments. 

What else can we say about the experimental results given the lessons we have learned from the modeling? Since the loops do not move relative to the fiber network, our modeling is applicable to adherent beads. If we were to consider nonadherent beads, however, then the fibers can move relative to the beads and collect in the interstitial places between the beads such that the beads become effectively larger and so percolate transversely to the compression at a smaller packing fraction than random-close packing~\cite{rcp} and perhaps even random-loose packing~\cite{rlp} given the uniaxial compression. In other words, if there is enough space for the fibers to move so that they do not have to bend around the beads, then there will be no compression stiffening.  It is interesting to note that with inert beads, the compression stiffening does not occur until a packing percentage of around 60\%~\cite{vanoosten2}, which is rather different from the nonadherent case discussed above. Interestingly, a different mechanism for compression stiffening that does not require bending but does involve a percolation of the area-conserving loops is possible (see Appendix B). The loops need not be rigid to drive the stiffening. 

To further test the notion of the bending of fibers as the driving compression stiffening mechanism, we replace fibrin with 2.4\% PAA gel, where bending is negligible. Here we do not observe compression stiffening even with 60\% dextran beads. See Fig. (\ref{PAA}). In fact, there may be a slight compression softening starting to occur around 20\% compressive strain.  This supports our finding that in the absence of bending, compression softening occurs due to the alignment of springs. For large enough strains however, we expect that the beads, held in place by the fiber network, eventually percolate transversely to the compression axis to lead to compression stiffening even in the absence of bending, as mentioned above.  

\section{Discussion}

For decades cells have been stretched, sheared, and compressed to understand their mechanics. We present a direct measurement of the compressive stress of a mouse embryonic fibroblastas as a function of compressive uniaxial strain imposed at the cell lengthscale and observe the nonlinear phenomenon of compression stiffening. The compression stiffening occurs around a compressive strain of 20\%. The implications of this finding are relevant at the cell scale and, potentially, at the tissue scale.  At the cell scale, we already know that cells stiffen when stretched and exhibit other nontrivial rheology~\cite{ott}.  Our compression stiffening result demonstrates that cells can behave nonlinearly under compression as well even on fast time scales where cytoskeletal reorganization is not feasible. Such behavior may indeed be important for cells in environments with intermediate to large homeostatic pressures. 

To interpret our compression stiffening finding at the cell scale, we study several different models. First, we consider the cell as a viscous medium modeled as an incompressible fluid surrounded by an actomyosin cortex modeled as a semiflexible loop and find compression stiffening. This model does not account for any fibers in the bulk of the cell, however, and a system-spanning bulk, rigid cytoskeletal fiber network would indeed contribute to a cell's mechanics. Since {\it in vitro} cytoskeletal filament networks compression soften, we have constructed a fiber network with fluid-based organelles and vesicles modeled as area-conserving loops randomly interspersed throughout the fiber network.  In the absence of any area-conserving loops, we find that the fiber network indeed compression softens due to the alignment of fibers along the axis perpendicular to the uniaxial compressive strain.  This new mechanism for compression softening is more collective than individual fiber buckling and demonstrates that softening can occur even in the absence of buckling. In the presence of area-conserving loops, we find that the network compression stiffens even for small packing fractions.  Not only do the area-conserving loops prevent the alignment of the fibers, they also promote the bending of the fibers to contort around them. A third mechanism for compression stiffening is due to angle-constraining crosslinks in the fiber network.  As the fiber network becomes increasingly compressed, the angles between fibers must distort resulting in an increasing stress in the network.  For this third mechanism, no area-conservation is required.

Of the three models, the one that best fits the data at least for up to 35\% compressive strain is the cell as a viscous medium enclosed by an actomyosin cortex. This model contains only one free parameter and suggests that the cortex of the three-dimensional cell subject to uniaxial compression can be viewed of as a set of loops forming a shell.  While interactions between the loops presumably exist, such interactions do not perhaps dominate the mechanics given our theoretical-experimental comparison using a single loop.   In addition, the bulk fiber network may be at play at larger strains and so we cannot rule out the fiber network modeling results. On the other hand, our freely-rotating crosslinked fiber network interspersed with area-conserving loops model can be more directly tested against {\it in vitro} fibrin networks embedded with dextran beads. Even with only 14\% packing percentage of beads, the fibrin network compression stiffens. Since the dextran beads are essentially rigid in the experiment, we rigidify the area-conserving loops by dramatically increasing the stiffness of the central-force springs surrounding the loops. While we find somewhat good agreement in the magnitude of the stress increase for a given strain range in the compression stiffening regime, the $\gamma_c$ is approximately 30\% for the experiment and just above 40\% in the modeling. We have made several speculations as to the difference in values between the experiment and the model bearing in mind that there already exists phenomenological modeling that does agree well the experimental data~\cite{vanoosten2}.  Our purpose here is to work with a more microscopic model with which we extract a new mechanism for the compression stiffening in terms of fiber bending. Interestingly, our results suggest that compression stiffening is robust for fiber networks in which the stretching energy of the filaments is comparable to the bending energy of filaments. This property is not applicable to actin crosslinked with, for example, fascin~\cite{gardel} or to PAA gels as evidenced by the lack of compression stiffening in such gels even with a high fraction of beads. This regime may be more accessible, however, with intermediate filaments such as vimentin and keratin~\cite{koster}. 

Our modeling also sheds light on the significant variation of cell stiffness measurements among different experimental experimental techniques and the choice of cell line~\cite{waterman,wirtz}.  If there is no spanning cytoskeletal network, then the moduli can be much lower than if there is a spanning cytoskeletal network present given the difference in changes in stress scale between the cortex surrounding a viscous medium model versus the fiber network with organelles model. Should the experimental method be more likely to probe the boundary of a cell as compared to its bulk, then different measurements may indeed be observed. Moreover, we find that the stress-strain curves are not only sensitive to boundary versus bulk measurements but, for the freely-rotating fiber network model, to the packing fraction and size distribution of area-conserving loops.  Such stress-strain curves could, therefore, provide a mechanical fingerprint to the size distribution of organelles in a cell.  Different cell-types have different size distributions and so one could distinguish between, say, an epithelial cell and a fibroblast given the stress-strain curve, in principle. As noted earlier, the compression of T-cells yields a cubic force-strain relationship up to strains about 30\%~\cite{lulevich}. As long as the force is proportional to the stress, our boundary cortex model in the limit of no-bending is relevant. T-cells have unusually large nuclei. Can we say something fundamental about the size of organelles such as the nucleus with respect to the size of the cell given our insights that go beyond the insights provided by Feric and Brangwynne for very large nuclei~\cite{feric}? For cells to exhibit larger compressive stresses, the presence of a bulk spanning fiber network is helpful. Perhaps for more migratory cells, the presence of a bulk spanning fiber network may hinder mobility in, say, confined environments.  Recent experiments find that vimentin-null mEFs migrate faster in microchannels then their wild-type counterparts~\cite{patteson2}.   

Now that we know individual cells compression stiffen, how do such nonlinear entities act when together in a compressed tissue? As stated in the introduction, one cannot directly imply that compression stiffening of tissues is caused by the compression stiffening of cells. Since, when we move across length scales, emergent phenomenon at a larger scale can exhibit behaviour otherwise unexpected from its constituents at a smaller scale. Yet, given that liver tissues almost completely lose their compression stiffening behaviour with decellularization \cite{perepelyuk}, it is plausible that one of the reasons of compression stiffening of tissues is indeed the compression stiffening of the individual building blocks. While phenomenological models~\cite{perepelyuk} and classical elasticity models~\cite{Engstrom19} approach compression stiffening directly from the tissue scale, our results here suggest that one can probe the tissue at smaller and smaller lengthscales to presumably find robustness of compression stiffening. At such scales, continuum mechanics may not be relevant, particularly for either extracellular matrix fibers and/or for cytoskeletal fibers. Tissue lacking in extracellular matrix is only as strong as its intercellular contacts.  While biology has presumably developed ways for cell-cell adhesion to depend on the nonlinearity of the cell's mechanics, an obvious answer presented here is to make a tissue composite where the area-conserving loops (or shells in three-dimensions) are now cells and the fibers are made of collagen. Given the ratio of stretching to bending moduli of individual collagen fibers~\cite{collagen}, a bundled network and/or one with angle-constraining crosslinks, will exhibit compression stiffening. Cells can also remodel the extracellular matrix on long enough time scales to make it more heterogeneous thereby adding to the complexity of the composite material. Indeed, biology has already mastered the highly nontrivial mechanics of compositeness in ways that we are just beginning to understand.

{\bf Conflicts of interests:} There are no conflicts of interest to declare. 

We thank an anonymous reviewer for queries which led to an analysis of the stretching and bending regimes in Sec. (\ref{One loop}). MCG acknowledges useful discussions with Matthias Merkel and Daniel Sussman, while JMS acknowledges useful exchanges with Dennis Discher and Tom Lubensky. KP acknowledges partial support from the National Science Center, Poland under Grant: UMO-2017/26/D/ST4/00997. AEP, AvO, and PAJ were supported by NSF DMR-1720530, NSF CMMI-154857 and NIH R01 EB017753. JMS acknowledges financial support from NSF-DMR-CMMT-1507938, NSF-DMR-CMMT-1832002, and from NSF-PHY-PoLS-1607416. 

\bibliography{References}

\begin{thebibliography}{53}%
\makeatletter
\providecommand \@ifxundefined [1]{%
 \@ifx{#1\undefined}
}%
\providecommand \@ifnum [1]{%
 \ifnum #1\expandafter \@firstoftwo
 \else \expandafter \@secondoftwo
 \fi
}%
\providecommand \@ifx [1]{%
 \ifx #1\expandafter \@firstoftwo
 \else \expandafter \@secondoftwo
 \fi
}%
\providecommand \natexlab [1]{#1}%
\providecommand \enquote  [1]{``#1''}%
\providecommand \bibnamefont  [1]{#1}%
\providecommand \bibfnamefont [1]{#1}%
\providecommand \citenamefont [1]{#1}%
\providecommand \href@noop [0]{\@secondoftwo}%
\providecommand \href [0]{\begingroup \@sanitize@url \@href}%
\providecommand \@href[1]{\@@startlink{#1}\@@href}%
\providecommand \@@href[1]{\endgroup#1\@@endlink}%
\providecommand \@sanitize@url [0]{\catcode `\\12\catcode `\$12\catcode
  `\&12\catcode `\#12\catcode `\^12\catcode `\_12\catcode `\%12\relax}%
\providecommand \@@startlink[1]{}%
\providecommand \@@endlink[0]{}%
\providecommand \url  [0]{\begingroup\@sanitize@url \@url }%
\providecommand \@url [1]{\endgroup\@href {#1}{\urlprefix }}%
\providecommand \urlprefix  [0]{URL }%
\providecommand \Eprint [0]{\href }%
\providecommand \doibase [0]{http://dx.doi.org/}%
\providecommand \selectlanguage [0]{\@gobble}%
\providecommand \bibinfo  [0]{\@secondoftwo}%
\providecommand \bibfield  [0]{\@secondoftwo}%
\providecommand \translation [1]{[#1]}%
\providecommand \BibitemOpen [0]{}%
\providecommand \bibitemStop [0]{}%
\providecommand \bibitemNoStop [0]{.\EOS\space}%
\providecommand \EOS [0]{\spacefactor3000\relax}%
\providecommand \BibitemShut  [1]{\csname bibitem#1\endcsname}%
\let\auto@bib@innerbib\@empty
\bibitem [{\citenamefont {Pogoda}\ \emph {et~al.}(2014)\citenamefont {Pogoda},
  \citenamefont {Chin}, \citenamefont {Georges}, \citenamefont {Byfield},
  \citenamefont {Bucki}, \citenamefont {Kim}, \citenamefont {Weaver},
  \citenamefont {Wells}, \citenamefont {Marcinkiewicz},\ and\ \citenamefont
  {Janmey}}]{pogoda}%
  \BibitemOpen
  \bibfield  {author} {\bibinfo {author} {\bibfnamefont {K.}~\bibnamefont
  {Pogoda}}, \bibinfo {author} {\bibfnamefont {L.}~\bibnamefont {Chin}},
  \bibinfo {author} {\bibfnamefont {P.~C.}\ \bibnamefont {Georges}}, \bibinfo
  {author} {\bibfnamefont {F.~J.}\ \bibnamefont {Byfield}}, \bibinfo {author}
  {\bibfnamefont {R.}~\bibnamefont {Bucki}}, \bibinfo {author} {\bibfnamefont
  {R.}~\bibnamefont {Kim}}, \bibinfo {author} {\bibfnamefont {M.}~\bibnamefont
  {Weaver}}, \bibinfo {author} {\bibfnamefont {R.~G.}\ \bibnamefont {Wells}},
  \bibinfo {author} {\bibfnamefont {C.}~\bibnamefont {Marcinkiewicz}}, \ and\
  \bibinfo {author} {\bibfnamefont {P.~A.}\ \bibnamefont {Janmey}},\
  }\href@noop {} {\bibfield  {journal} {\bibinfo  {journal} {New J. Phys.}\
  }\textbf {\bibinfo {volume} {16}},\ \bibinfo {pages} {075002} (\bibinfo
  {year} {2014})}\BibitemShut {NoStop}%
\bibitem [{\citenamefont {Perepelyuk}\ \emph {et~al.}(2016)\citenamefont
  {Perepelyuk}, \citenamefont {Chin}, \citenamefont {Cao}, \citenamefont {van
  Oosten}, \citenamefont {Shenoy}, \citenamefont {Janmey},\ and\ \citenamefont
  {Wells}}]{perepelyuk}%
  \BibitemOpen
  \bibfield  {author} {\bibinfo {author} {\bibfnamefont {M.}~\bibnamefont
  {Perepelyuk}}, \bibinfo {author} {\bibfnamefont {L.}~\bibnamefont {Chin}},
  \bibinfo {author} {\bibfnamefont {X.}~\bibnamefont {Cao}}, \bibinfo {author}
  {\bibfnamefont {A.}~\bibnamefont {van Oosten}}, \bibinfo {author}
  {\bibfnamefont {V.~B.}\ \bibnamefont {Shenoy}}, \bibinfo {author}
  {\bibfnamefont {P.~A.}\ \bibnamefont {Janmey}}, \ and\ \bibinfo {author}
  {\bibfnamefont {R.~G.}\ \bibnamefont {Wells}},\ }\href@noop {} {\bibfield
  {journal} {\bibinfo  {journal} {PloS ONE}\ }\textbf {\bibinfo {volume}
  {11}},\ \bibinfo {pages} {e0146588} (\bibinfo {year} {2016})}\BibitemShut
  {NoStop}%
\bibitem [{\citenamefont {Engstrom}\ \emph {et~al.}(2019)\citenamefont
  {Engstrom}, \citenamefont {Pogoda}, \citenamefont {Cruz}, \citenamefont
  {Janmey},\ and\ \citenamefont {Schwarz}}]{Engstrom19}%
  \BibitemOpen
  \bibfield  {author} {\bibinfo {author} {\bibfnamefont {T.~A.}\ \bibnamefont
  {Engstrom}}, \bibinfo {author} {\bibfnamefont {K.}~\bibnamefont {Pogoda}},
  \bibinfo {author} {\bibfnamefont {K.}~\bibnamefont {Cruz}}, \bibinfo {author}
  {\bibfnamefont {P.~A.}\ \bibnamefont {Janmey}}, \ and\ \bibinfo {author}
  {\bibfnamefont {J.~M.}\ \bibnamefont {Schwarz}},\ }\href@noop {} {\bibfield
  {journal} {\bibinfo  {journal} {Phys. Rev. E}\ }\textbf {\bibinfo {volume}
  {99}},\ \bibinfo {pages} {052413} (\bibinfo {year} {2019})}\BibitemShut
  {NoStop}%
\bibitem [{\citenamefont {Alberts}\ \emph {et~al.}(1994)\citenamefont
  {Alberts}, \citenamefont {Bray}, \citenamefont {Lewis}, \citenamefont {Raff},
  \citenamefont {Roberts},\ and\ \citenamefont {Watson}}]{alberts}%
  \BibitemOpen
  \bibfield  {author} {\bibinfo {author} {\bibfnamefont {B.}~\bibnamefont
  {Alberts}}, \bibinfo {author} {\bibfnamefont {D.}~\bibnamefont {Bray}},
  \bibinfo {author} {\bibfnamefont {J.}~\bibnamefont {Lewis}}, \bibinfo
  {author} {\bibfnamefont {M.}~\bibnamefont {Raff}}, \bibinfo {author}
  {\bibfnamefont {K.}~\bibnamefont {Roberts}}, \ and\ \bibinfo {author}
  {\bibfnamefont {J.}~\bibnamefont {Watson}},\ }\href@noop {} {\emph {\bibinfo
  {title} {Molecular Biology of the Cell}}}\ (\bibinfo  {publisher} {Garland,
  New York},\ \bibinfo {year} {1994})\ pp.\ \bibinfo {pages}
  {907--982}\BibitemShut {NoStop}%
\bibitem [{\citenamefont {Gardel}\ \emph {et~al.}(2004)\citenamefont {Gardel},
  \citenamefont {Shin}, \citenamefont {MacKintosh}, \citenamefont {Mahadevan},
  \citenamefont {Matsudaira},\ and\ \citenamefont {Weitz}}]{gardel}%
  \BibitemOpen
  \bibfield  {author} {\bibinfo {author} {\bibfnamefont {M.}~\bibnamefont
  {Gardel}}, \bibinfo {author} {\bibfnamefont {J.}~\bibnamefont {Shin}},
  \bibinfo {author} {\bibfnamefont {F.}~\bibnamefont {MacKintosh}}, \bibinfo
  {author} {\bibfnamefont {L.}~\bibnamefont {Mahadevan}}, \bibinfo {author}
  {\bibfnamefont {P.}~\bibnamefont {Matsudaira}}, \ and\ \bibinfo {author}
  {\bibfnamefont {D.}~\bibnamefont {Weitz}},\ }\href@noop {} {\bibfield
  {journal} {\bibinfo  {journal} {Science}\ }\textbf {\bibinfo {volume}
  {304}},\ \bibinfo {pages} {1301} (\bibinfo {year} {2004})}\BibitemShut
  {NoStop}%
\bibitem [{\citenamefont {M{\"u}cke}\ \emph {et~al.}(2004)\citenamefont
  {M{\"u}cke}, \citenamefont {Kreplak}, \citenamefont {Kirmse}, \citenamefont
  {Wedig}, \citenamefont {Herrmann}, \citenamefont {Aebi},\ and\ \citenamefont
  {Langowski}}]{mucke}%
  \BibitemOpen
  \bibfield  {author} {\bibinfo {author} {\bibfnamefont {N.}~\bibnamefont
  {M{\"u}cke}}, \bibinfo {author} {\bibfnamefont {L.}~\bibnamefont {Kreplak}},
  \bibinfo {author} {\bibfnamefont {R.}~\bibnamefont {Kirmse}}, \bibinfo
  {author} {\bibfnamefont {T.}~\bibnamefont {Wedig}}, \bibinfo {author}
  {\bibfnamefont {H.}~\bibnamefont {Herrmann}}, \bibinfo {author}
  {\bibfnamefont {U.}~\bibnamefont {Aebi}}, \ and\ \bibinfo {author}
  {\bibfnamefont {J.}~\bibnamefont {Langowski}},\ }\href@noop {} {\bibfield
  {journal} {\bibinfo  {journal} {J. Mol. Biol.}\ }\textbf {\bibinfo {volume}
  {335}},\ \bibinfo {pages} {1241} (\bibinfo {year} {2004})}\BibitemShut
  {NoStop}%
\bibitem [{\citenamefont {Gittes}\ \emph {et~al.}(1993)\citenamefont {Gittes},
  \citenamefont {Mickey}, \citenamefont {Nettleton},\ and\ \citenamefont
  {Howard}}]{gittes1993}%
  \BibitemOpen
  \bibfield  {author} {\bibinfo {author} {\bibfnamefont {F.}~\bibnamefont
  {Gittes}}, \bibinfo {author} {\bibfnamefont {B.}~\bibnamefont {Mickey}},
  \bibinfo {author} {\bibfnamefont {J.}~\bibnamefont {Nettleton}}, \ and\
  \bibinfo {author} {\bibfnamefont {J.}~\bibnamefont {Howard}},\ }\href@noop {}
  {\bibfield  {journal} {\bibinfo  {journal} {J. Cell Biol.}\ }\textbf
  {\bibinfo {volume} {120}},\ \bibinfo {pages} {923} (\bibinfo {year}
  {1993})}\BibitemShut {NoStop}%
\bibitem [{\citenamefont {Ott}\ \emph {et~al.}(1993)\citenamefont {Ott},
  \citenamefont {Magnasco}, \citenamefont {Simon},\ and\ \citenamefont
  {Libchaber}}]{ott1993}%
  \BibitemOpen
  \bibfield  {author} {\bibinfo {author} {\bibfnamefont {A.}~\bibnamefont
  {Ott}}, \bibinfo {author} {\bibfnamefont {M.}~\bibnamefont {Magnasco}},
  \bibinfo {author} {\bibfnamefont {A.}~\bibnamefont {Simon}}, \ and\ \bibinfo
  {author} {\bibfnamefont {A.}~\bibnamefont {Libchaber}},\ }\href@noop {}
  {\bibfield  {journal} {\bibinfo  {journal} {Phys. Rev. E}\ }\textbf {\bibinfo
  {volume} {48}},\ \bibinfo {pages} {R1642} (\bibinfo {year}
  {1993})}\BibitemShut {NoStop}%
\bibitem [{\citenamefont {Shabbir}\ \emph {et~al.}(2014)\citenamefont
  {Shabbir}, \citenamefont {Cleland}, \citenamefont {Goldman},\ and\
  \citenamefont {Mrksich}}]{shabbir2014geometric}%
  \BibitemOpen
  \bibfield  {author} {\bibinfo {author} {\bibfnamefont {S.~H.}\ \bibnamefont
  {Shabbir}}, \bibinfo {author} {\bibfnamefont {M.~M.}\ \bibnamefont
  {Cleland}}, \bibinfo {author} {\bibfnamefont {R.~D.}\ \bibnamefont
  {Goldman}}, \ and\ \bibinfo {author} {\bibfnamefont {M.}~\bibnamefont
  {Mrksich}},\ }\href@noop {} {\bibfield  {journal} {\bibinfo  {journal}
  {Biomaterials}\ }\textbf {\bibinfo {volume} {35}},\ \bibinfo {pages} {1359}
  (\bibinfo {year} {2014})}\BibitemShut {NoStop}%
\bibitem [{\citenamefont {Guo}\ \emph {et~al.}(2013)\citenamefont {Guo},
  \citenamefont {Ehrlicher}, \citenamefont {Mahammad}, \citenamefont {Fabich},
  \citenamefont {Jensen}, \citenamefont {Moore}, \citenamefont {Fredberg},
  \citenamefont {Goldman},\ and\ \citenamefont {Weitz}}]{guo2013role}%
  \BibitemOpen
  \bibfield  {author} {\bibinfo {author} {\bibfnamefont {M.}~\bibnamefont
  {Guo}}, \bibinfo {author} {\bibfnamefont {A.~J.}\ \bibnamefont {Ehrlicher}},
  \bibinfo {author} {\bibfnamefont {S.}~\bibnamefont {Mahammad}}, \bibinfo
  {author} {\bibfnamefont {H.}~\bibnamefont {Fabich}}, \bibinfo {author}
  {\bibfnamefont {M.~H.}\ \bibnamefont {Jensen}}, \bibinfo {author}
  {\bibfnamefont {J.~R.}\ \bibnamefont {Moore}}, \bibinfo {author}
  {\bibfnamefont {J.~J.}\ \bibnamefont {Fredberg}}, \bibinfo {author}
  {\bibfnamefont {R.~D.}\ \bibnamefont {Goldman}}, \ and\ \bibinfo {author}
  {\bibfnamefont {D.~A.}\ \bibnamefont {Weitz}},\ }\href@noop {} {\bibfield
  {journal} {\bibinfo  {journal} {Biophy. J.}\ }\textbf {\bibinfo {volume}
  {105}},\ \bibinfo {pages} {1562} (\bibinfo {year} {2013})}\BibitemShut
  {NoStop}%
\bibitem [{\citenamefont {Mendez}\ \emph {et~al.}(2014)\citenamefont {Mendez},
  \citenamefont {Restle},\ and\ \citenamefont {Janmey}}]{Melissa}%
  \BibitemOpen
  \bibfield  {author} {\bibinfo {author} {\bibfnamefont {M.~G.}\ \bibnamefont
  {Mendez}}, \bibinfo {author} {\bibfnamefont {D.}~\bibnamefont {Restle}}, \
  and\ \bibinfo {author} {\bibfnamefont {P.~A.}\ \bibnamefont {Janmey}},\
  }\href@noop {} {\bibfield  {journal} {\bibinfo  {journal} {Biophys. J.}\
  }\textbf {\bibinfo {volume} {107}},\ \bibinfo {pages} {314} (\bibinfo {year}
  {2014})}\BibitemShut {NoStop}%
\bibitem [{\citenamefont {Patteson}\ \emph
  {et~al.}(2019{\natexlab{a}})\citenamefont {Patteson}, \citenamefont
  {Vahabikashi}, \citenamefont {Pogoda}, \citenamefont {Adam}, \citenamefont
  {Mandal}, \citenamefont {Kittisopikul}, \citenamefont {Sivagurunathan},
  \citenamefont {Goldman}, \citenamefont {Goldman},\ and\ \citenamefont
  {Janmey}}]{Patteson1}%
  \BibitemOpen
  \bibfield  {author} {\bibinfo {author} {\bibfnamefont {A.~E.}\ \bibnamefont
  {Patteson}}, \bibinfo {author} {\bibfnamefont {A.}~\bibnamefont
  {Vahabikashi}}, \bibinfo {author} {\bibfnamefont {K.}~\bibnamefont {Pogoda}},
  \bibinfo {author} {\bibfnamefont {S.~A.}\ \bibnamefont {Adam}}, \bibinfo
  {author} {\bibfnamefont {K.}~\bibnamefont {Mandal}}, \bibinfo {author}
  {\bibfnamefont {M.}~\bibnamefont {Kittisopikul}}, \bibinfo {author}
  {\bibfnamefont {S.}~\bibnamefont {Sivagurunathan}}, \bibinfo {author}
  {\bibfnamefont {A.}~\bibnamefont {Goldman}}, \bibinfo {author} {\bibfnamefont
  {R.~D.}\ \bibnamefont {Goldman}}, \ and\ \bibinfo {author} {\bibfnamefont
  {P.~A.}\ \bibnamefont {Janmey}},\ }\href@noop {} {\bibfield  {journal}
  {\bibinfo  {journal} {Journal of Cell Biology}\ }\textbf {\bibinfo {volume}
  {218}},\ \bibinfo {pages} {4079} (\bibinfo {year}
  {2019}{\natexlab{a}})}\BibitemShut {NoStop}%
\bibitem [{\citenamefont {Janmey}\ \emph {et~al.}(2006)\citenamefont {Janmey},
  \citenamefont {McCormick}, \citenamefont {Rammensee}, \citenamefont {Leight},
  \citenamefont {Georges},\ and\ \citenamefont {MacKintosh}}]{janmey}%
  \BibitemOpen
  \bibfield  {author} {\bibinfo {author} {\bibfnamefont {P.~A.}\ \bibnamefont
  {Janmey}}, \bibinfo {author} {\bibfnamefont {M.~E.}\ \bibnamefont
  {McCormick}}, \bibinfo {author} {\bibfnamefont {S.}~\bibnamefont
  {Rammensee}}, \bibinfo {author} {\bibfnamefont {J.~L.}\ \bibnamefont
  {Leight}}, \bibinfo {author} {\bibfnamefont {P.~C.}\ \bibnamefont {Georges}},
  \ and\ \bibinfo {author} {\bibfnamefont {F.~C.}\ \bibnamefont {MacKintosh}},\
  }\href@noop {} {\bibfield  {journal} {\bibinfo  {journal} {Nat. Mat.}\
  }\textbf {\bibinfo {volume} {6}},\ \bibinfo {pages} {48} (\bibinfo {year}
  {2006})}\BibitemShut {NoStop}%
\bibitem [{\citenamefont {Storm}\ \emph {et~al.}(2005)\citenamefont {Storm},
  \citenamefont {Pastore}, \citenamefont {MacKintosh}, \citenamefont
  {Lubensky},\ and\ \citenamefont {Janmey}}]{storm}%
  \BibitemOpen
  \bibfield  {author} {\bibinfo {author} {\bibfnamefont {C.}~\bibnamefont
  {Storm}}, \bibinfo {author} {\bibfnamefont {J.~J.}\ \bibnamefont {Pastore}},
  \bibinfo {author} {\bibfnamefont {F.~C.}\ \bibnamefont {MacKintosh}},
  \bibinfo {author} {\bibfnamefont {T.~C.}\ \bibnamefont {Lubensky}}, \ and\
  \bibinfo {author} {\bibfnamefont {P.~A.}\ \bibnamefont {Janmey}},\
  }\href@noop {} {\bibfield  {journal} {\bibinfo  {journal} {Nature}\ }\textbf
  {\bibinfo {volume} {435}},\ \bibinfo {pages} {191} (\bibinfo {year}
  {2005})}\BibitemShut {NoStop}%
\bibitem [{\citenamefont {Van~Oosten}\ \emph {et~al.}(2016)\citenamefont
  {Van~Oosten}, \citenamefont {Vahabi}, \citenamefont {Licup}, \citenamefont
  {Sharma}, \citenamefont {Galie}, \citenamefont {MacKintosh},\ and\
  \citenamefont {Janmey}}]{vanoosten}%
  \BibitemOpen
  \bibfield  {author} {\bibinfo {author} {\bibfnamefont {A.~S.}\ \bibnamefont
  {Van~Oosten}}, \bibinfo {author} {\bibfnamefont {M.}~\bibnamefont {Vahabi}},
  \bibinfo {author} {\bibfnamefont {A.~J.}\ \bibnamefont {Licup}}, \bibinfo
  {author} {\bibfnamefont {A.}~\bibnamefont {Sharma}}, \bibinfo {author}
  {\bibfnamefont {P.~A.}\ \bibnamefont {Galie}}, \bibinfo {author}
  {\bibfnamefont {F.~C.}\ \bibnamefont {MacKintosh}}, \ and\ \bibinfo {author}
  {\bibfnamefont {P.~A.}\ \bibnamefont {Janmey}},\ }\href@noop {} {\bibfield
  {journal} {\bibinfo  {journal} {Sci. Reps.}\ }\textbf {\bibinfo {volume}
  {6}},\ \bibinfo {pages} {19270} (\bibinfo {year} {2016})}\BibitemShut
  {NoStop}%
\bibitem [{\citenamefont {MacKintosh}\ \emph {et~al.}(1995)\citenamefont
  {MacKintosh}, \citenamefont {K{\"a}s},\ and\ \citenamefont
  {Janmey}}]{mackintosh1995elasticity}%
  \BibitemOpen
  \bibfield  {author} {\bibinfo {author} {\bibfnamefont {F.}~\bibnamefont
  {MacKintosh}}, \bibinfo {author} {\bibfnamefont {J.}~\bibnamefont {K{\"a}s}},
  \ and\ \bibinfo {author} {\bibfnamefont {P.}~\bibnamefont {Janmey}},\
  }\href@noop {} {\bibfield  {journal} {\bibinfo  {journal} {Phys. Rev. Lett.}\
  }\textbf {\bibinfo {volume} {75}},\ \bibinfo {pages} {4425} (\bibinfo {year}
  {1995})}\BibitemShut {NoStop}%
\bibitem [{\citenamefont {Broedersz}\ and\ \citenamefont
  {MacKintosh}(2014)}]{broedersz}%
  \BibitemOpen
  \bibfield  {author} {\bibinfo {author} {\bibfnamefont {C.~P.}\ \bibnamefont
  {Broedersz}}\ and\ \bibinfo {author} {\bibfnamefont {F.~C.}\ \bibnamefont
  {MacKintosh}},\ }\href@noop {} {\bibfield  {journal} {\bibinfo  {journal}
  {Rev. Mod. Phys.}\ }\textbf {\bibinfo {volume} {86}},\ \bibinfo {pages} {995}
  (\bibinfo {year} {2014})}\BibitemShut {NoStop}%
\bibitem [{\citenamefont {Landau}\ and\ \citenamefont
  {Lifshitz}(1986)}]{landau}%
  \BibitemOpen
  \bibfield  {author} {\bibinfo {author} {\bibfnamefont {L.~D.}\ \bibnamefont
  {Landau}}\ and\ \bibinfo {author} {\bibfnamefont {E.}~\bibnamefont
  {Lifshitz}},\ }\href@noop {} {\bibfield  {journal} {\bibinfo  {journal}
  {Course of Theoretical Physics}\ }\textbf {\bibinfo {volume} {3}},\ \bibinfo
  {pages} {109} (\bibinfo {year} {1986})}\BibitemShut {NoStop}%
\bibitem [{\citenamefont {Pilyugina}\ \emph {et~al.}(2017)\citenamefont
  {Pilyugina}, \citenamefont {Krajina}, \citenamefont {Spakowitz},\ and\
  \citenamefont {Scheiber}}]{pilyugina}%
  \BibitemOpen
  \bibfield  {author} {\bibinfo {author} {\bibfnamefont {E.}~\bibnamefont
  {Pilyugina}}, \bibinfo {author} {\bibfnamefont {B.}~\bibnamefont {Krajina}},
  \bibinfo {author} {\bibfnamefont {A.~J.}\ \bibnamefont {Spakowitz}}, \ and\
  \bibinfo {author} {\bibfnamefont {J.~D.}\ \bibnamefont {Scheiber}},\
  }\href@noop {} {\bibfield  {journal} {\bibinfo  {journal} {Polymers}\
  }\textbf {\bibinfo {volume} {9}},\ \bibinfo {pages} {99} (\bibinfo {year}
  {2017})}\BibitemShut {NoStop}%
\bibitem [{\citenamefont {Das}\ \emph {et~al.}(2012)\citenamefont {Das},
  \citenamefont {Quint},\ and\ \citenamefont {Schwarz}}]{das}%
  \BibitemOpen
  \bibfield  {author} {\bibinfo {author} {\bibfnamefont {M.}~\bibnamefont
  {Das}}, \bibinfo {author} {\bibfnamefont {D.}~\bibnamefont {Quint}}, \ and\
  \bibinfo {author} {\bibfnamefont {J.}~\bibnamefont {Schwarz}},\ }\href@noop
  {} {\bibfield  {journal} {\bibinfo  {journal} {PloS ONE}\ }\textbf {\bibinfo
  {volume} {7}},\ \bibinfo {pages} {e35939} (\bibinfo {year}
  {2012})}\BibitemShut {NoStop}%
\bibitem [{\citenamefont {Lin}\ and\ \citenamefont {Gu}(2015)}]{lin}%
  \BibitemOpen
  \bibfield  {author} {\bibinfo {author} {\bibfnamefont {S.}~\bibnamefont
  {Lin}}\ and\ \bibinfo {author} {\bibfnamefont {L.}~\bibnamefont {Gu}},\
  }\href@noop {} {\bibfield  {journal} {\bibinfo  {journal} {Materials
  (Basel)}\ }\textbf {\bibinfo {volume} {8}},\ \bibinfo {pages} {551} (\bibinfo
  {year} {2015})}\BibitemShut {NoStop}%
\bibitem [{\citenamefont {Hatami-Marbini}(2018)}]{hatami}%
  \BibitemOpen
  \bibfield  {author} {\bibinfo {author} {\bibfnamefont {H.}~\bibnamefont
  {Hatami-Marbini}},\ }\href@noop {} {\bibfield  {journal} {\bibinfo  {journal}
  {Phys. Rev. E}\ }\textbf {\bibinfo {volume} {97}},\ \bibinfo {pages} {022504}
  (\bibinfo {year} {2018})}\BibitemShut {NoStop}%
\bibitem [{\citenamefont {Picariello}\ and\ \citenamefont
  {et~al.}(2019)}]{hannah}%
  \BibitemOpen
  \bibfield  {author} {\bibinfo {author} {\bibfnamefont {H.}~\bibnamefont
  {Picariello}}\ and\ \bibinfo {author} {\bibnamefont {et~al.}},\ }\href@noop
  {} {\bibfield  {journal} {\bibinfo  {journal} {PNAS}\ ,\ \bibinfo {pages}
  {201902847}} (\bibinfo {year} {2019})}\BibitemShut {NoStop}%
\bibitem [{\citenamefont {van Oosten}\ \emph {et~al.}(2019)\citenamefont {van
  Oosten}, \citenamefont {Chen}, \citenamefont {Chin}, \citenamefont {Cruz},
  \citenamefont {Patteson}, \citenamefont {Pogoda}, \citenamefont {Shenoy},\
  and\ \citenamefont {Janmey}}]{vanoosten2}%
  \BibitemOpen
  \bibfield  {author} {\bibinfo {author} {\bibfnamefont {A.~S.}\ \bibnamefont
  {van Oosten}}, \bibinfo {author} {\bibfnamefont {X.}~\bibnamefont {Chen}},
  \bibinfo {author} {\bibfnamefont {L.}~\bibnamefont {Chin}}, \bibinfo {author}
  {\bibfnamefont {K.}~\bibnamefont {Cruz}}, \bibinfo {author} {\bibfnamefont
  {A.~E.}\ \bibnamefont {Patteson}}, \bibinfo {author} {\bibfnamefont
  {K.}~\bibnamefont {Pogoda}}, \bibinfo {author} {\bibfnamefont {V.~B.}\
  \bibnamefont {Shenoy}}, \ and\ \bibinfo {author} {\bibfnamefont {P.~A.}\
  \bibnamefont {Janmey}},\ }\href@noop {} {\bibfield  {journal} {\bibinfo
  {journal} {Nature}\ }\textbf {\bibinfo {volume} {573}},\ \bibinfo {pages}
  {96} (\bibinfo {year} {2019})}\BibitemShut {NoStop}%
\bibitem [{\citenamefont {Onoda}\ and\ \citenamefont {Liniger}(1990)}]{rlp}%
  \BibitemOpen
  \bibfield  {author} {\bibinfo {author} {\bibfnamefont {G.~Y.}\ \bibnamefont
  {Onoda}}\ and\ \bibinfo {author} {\bibfnamefont {E.~G.}\ \bibnamefont
  {Liniger}},\ }\href@noop {} {\bibfield  {journal} {\bibinfo  {journal} {Phys.
  Rev. Lett.}\ }\textbf {\bibinfo {volume} {64}},\ \bibinfo {pages} {2727}
  (\bibinfo {year} {1990})}\BibitemShut {NoStop}%
\bibitem [{\citenamefont {Bernal}\ and\ \citenamefont {Mason}(1960)}]{rcp}%
  \BibitemOpen
  \bibfield  {author} {\bibinfo {author} {\bibfnamefont {J.}~\bibnamefont
  {Bernal}}\ and\ \bibinfo {author} {\bibfnamefont {J.}~\bibnamefont {Mason}},\
  }\href@noop {} {\bibfield  {journal} {\bibinfo  {journal} {Nature}\ }\textbf
  {\bibinfo {volume} {188}},\ \bibinfo {pages} {910} (\bibinfo {year}
  {1960})}\BibitemShut {NoStop}%
\bibitem [{\citenamefont {Maxwell}(1864)}]{maxwell}%
  \BibitemOpen
  \bibfield  {author} {\bibinfo {author} {\bibfnamefont {J.~C.}\ \bibnamefont
  {Maxwell}},\ }\href@noop {} {\bibfield  {journal} {\bibinfo  {journal}
  {Philos. Mag. Ser. 5}\ }\textbf {\bibinfo {volume} {27}},\ \bibinfo {pages}
  {294} (\bibinfo {year} {1864})}\BibitemShut {NoStop}%
\bibitem [{\citenamefont {Lulevich}\ \emph {et~al.}(2006)\citenamefont
  {Lulevich}, \citenamefont {Zink}, \citenamefont {Chen}, \citenamefont {Liu},\
  and\ \citenamefont {Liu}}]{lulevich}%
  \BibitemOpen
  \bibfield  {author} {\bibinfo {author} {\bibfnamefont {V.}~\bibnamefont
  {Lulevich}}, \bibinfo {author} {\bibfnamefont {T.}~\bibnamefont {Zink}},
  \bibinfo {author} {\bibfnamefont {H.-Y.}\ \bibnamefont {Chen}}, \bibinfo
  {author} {\bibfnamefont {F.-T.}\ \bibnamefont {Liu}}, \ and\ \bibinfo
  {author} {\bibfnamefont {G.-Y.}\ \bibnamefont {Liu}},\ }\href@noop {}
  {\bibfield  {journal} {\bibinfo  {journal} {Langmuir}\ }\textbf {\bibinfo
  {volume} {22}},\ \bibinfo {pages} {8151} (\bibinfo {year}
  {2006})}\BibitemShut {NoStop}%
\bibitem [{\citenamefont {Arreaga}\ \emph {et~al.}(2002)\citenamefont
  {Arreaga}, \citenamefont {Capovilla}, \citenamefont {Chryssomalakos},\ and\
  \citenamefont {Guven}}]{arreaga}%
  \BibitemOpen
  \bibfield  {author} {\bibinfo {author} {\bibfnamefont {G.}~\bibnamefont
  {Arreaga}}, \bibinfo {author} {\bibfnamefont {R.}~\bibnamefont {Capovilla}},
  \bibinfo {author} {\bibfnamefont {C.}~\bibnamefont {Chryssomalakos}}, \ and\
  \bibinfo {author} {\bibfnamefont {J.}~\bibnamefont {Guven}},\ }\href@noop {}
  {\bibfield  {journal} {\bibinfo  {journal} {Phys. Rev. E}\ }\textbf {\bibinfo
  {volume} {65}},\ \bibinfo {pages} {031801} (\bibinfo {year}
  {2002})}\BibitemShut {NoStop}%
\bibitem [{\citenamefont {Hiramoto}(1963)}]{hiramoto1963}%
  \BibitemOpen
  \bibfield  {author} {\bibinfo {author} {\bibfnamefont {Y.}~\bibnamefont
  {Hiramoto}},\ }\href@noop {} {\bibfield  {journal} {\bibinfo  {journal} {Exp.
  Cell Res.}\ }\textbf {\bibinfo {volume} {32}},\ \bibinfo {pages} {59}
  (\bibinfo {year} {1963})}\BibitemShut {NoStop}%
\bibitem [{\citenamefont {Feng}\ \emph {et~al.}(2016)\citenamefont {Feng},
  \citenamefont {Levine}, \citenamefont {Mao},\ and\ \citenamefont
  {Sander}}]{feng2016nonlinear}%
  \BibitemOpen
  \bibfield  {author} {\bibinfo {author} {\bibfnamefont {J.}~\bibnamefont
  {Feng}}, \bibinfo {author} {\bibfnamefont {H.}~\bibnamefont {Levine}},
  \bibinfo {author} {\bibfnamefont {X.}~\bibnamefont {Mao}}, \ and\ \bibinfo
  {author} {\bibfnamefont {L.~M.}\ \bibnamefont {Sander}},\ }\href@noop {}
  {\bibfield  {journal} {\bibinfo  {journal} {Soft matter}\ }\textbf {\bibinfo
  {volume} {12}},\ \bibinfo {pages} {1419} (\bibinfo {year}
  {2016})}\BibitemShut {NoStop}%
\bibitem [{\citenamefont {Fletcher}\ and\ \citenamefont
  {Mullins}(2010)}]{fibercouplings}%
  \BibitemOpen
  \bibfield  {author} {\bibinfo {author} {\bibfnamefont {D.~A.}\ \bibnamefont
  {Fletcher}}\ and\ \bibinfo {author} {\bibfnamefont {R.~D.}\ \bibnamefont
  {Mullins}},\ }\href@noop {} {\bibfield  {journal} {\bibinfo  {journal}
  {Nature}\ }\textbf {\bibinfo {volume} {463}},\ \bibinfo {pages} {485}
  (\bibinfo {year} {2010})}\BibitemShut {NoStop}%
\bibitem [{\citenamefont {Vahabi}\ \emph {et~al.}(2016)\citenamefont {Vahabi},
  \citenamefont {Sharma}, \citenamefont {Licup}, \citenamefont {van Oosten},
  \citenamefont {Galie}, \citenamefont {Janmey},\ and\ \citenamefont
  {MacKintosh}}]{prestress}%
  \BibitemOpen
  \bibfield  {author} {\bibinfo {author} {\bibfnamefont {M.}~\bibnamefont
  {Vahabi}}, \bibinfo {author} {\bibfnamefont {A.}~\bibnamefont {Sharma}},
  \bibinfo {author} {\bibfnamefont {A.~J.}\ \bibnamefont {Licup}}, \bibinfo
  {author} {\bibfnamefont {A.~S.~G.}\ \bibnamefont {van Oosten}}, \bibinfo
  {author} {\bibfnamefont {P.~A.}\ \bibnamefont {Galie}}, \bibinfo {author}
  {\bibfnamefont {P.~A.}\ \bibnamefont {Janmey}}, \ and\ \bibinfo {author}
  {\bibfnamefont {F.~C.}\ \bibnamefont {MacKintosh}},\ }\href@noop {}
  {\bibfield  {journal} {\bibinfo  {journal} {Soft Matt.}\ }\textbf {\bibinfo
  {volume} {12}},\ \bibinfo {pages} {5050} (\bibinfo {year}
  {2016})}\BibitemShut {NoStop}%
\bibitem [{\citenamefont {Charrier}\ and\ \citenamefont
  {Janmey}(2016)}]{charrier}%
  \BibitemOpen
  \bibfield  {author} {\bibinfo {author} {\bibfnamefont {E.~E.}\ \bibnamefont
  {Charrier}}\ and\ \bibinfo {author} {\bibfnamefont {P.~A.}\ \bibnamefont
  {Janmey}},\ }\href@noop {} {\bibfield  {journal} {\bibinfo  {journal} {Meth.
  Enzymol.}\ }\textbf {\bibinfo {volume} {568}},\ \bibinfo {pages} {35}
  (\bibinfo {year} {2016})}\BibitemShut {NoStop}%
\bibitem [{\citenamefont {Discher}\ \emph {et~al.}(1997)\citenamefont
  {Discher}, \citenamefont {Boal},\ and\ \citenamefont {Boey}}]{discher1997}%
  \BibitemOpen
  \bibfield  {author} {\bibinfo {author} {\bibfnamefont {D.~E.}\ \bibnamefont
  {Discher}}, \bibinfo {author} {\bibfnamefont {D.~H.}\ \bibnamefont {Boal}}, \
  and\ \bibinfo {author} {\bibfnamefont {S.~K.}\ \bibnamefont {Boey}},\
  }\href@noop {} {\bibfield  {journal} {\bibinfo  {journal} {Phys. Rev. E}\
  }\textbf {\bibinfo {volume} {55}},\ \bibinfo {pages} {4762} (\bibinfo {year}
  {1997})}\BibitemShut {NoStop}%
\bibitem [{\citenamefont {Wintz}\ \emph {et~al.}(1997)\citenamefont {Wintz},
  \citenamefont {Everaers},\ and\ \citenamefont {Seifert}}]{wintz1997}%
  \BibitemOpen
  \bibfield  {author} {\bibinfo {author} {\bibfnamefont {W.}~\bibnamefont
  {Wintz}}, \bibinfo {author} {\bibfnamefont {R.}~\bibnamefont {Everaers}}, \
  and\ \bibinfo {author} {\bibfnamefont {U.}~\bibnamefont {Seifert}},\
  }\href@noop {} {\bibfield  {journal} {\bibinfo  {journal} {Journal de
  Physique I}\ }\textbf {\bibinfo {volume} {7}},\ \bibinfo {pages} {1097}
  (\bibinfo {year} {1997})}\BibitemShut {NoStop}%
\bibitem [{\citenamefont {Beysens}\ and\ \citenamefont
  {Forgacs}(2013)}]{beysens2013dynamical}%
  \BibitemOpen
  \bibfield  {author} {\bibinfo {author} {\bibfnamefont {D.}~\bibnamefont
  {Beysens}}\ and\ \bibinfo {author} {\bibfnamefont {G.}~\bibnamefont
  {Forgacs}},\ }\href@noop {} {\emph {\bibinfo {title} {Dynamical Networks in
  Physics and Biology: At the Frontier of Physics and Biology Les Houches
  Workshop, March 17--21, 1997}}},\ Vol.~\bibinfo {volume} {10}\ (\bibinfo
  {publisher} {Springer Science \& Business Media},\ \bibinfo {year}
  {2013})\BibitemShut {NoStop}%
\bibitem [{\citenamefont {Head}\ \emph {et~al.}(2003)\citenamefont {Head},
  \citenamefont {Levine},\ and\ \citenamefont
  {MacKintosh}}]{head2003deformation}%
  \BibitemOpen
  \bibfield  {author} {\bibinfo {author} {\bibfnamefont {D.~A.}\ \bibnamefont
  {Head}}, \bibinfo {author} {\bibfnamefont {A.~J.}\ \bibnamefont {Levine}}, \
  and\ \bibinfo {author} {\bibfnamefont {F.}~\bibnamefont {MacKintosh}},\
  }\href@noop {} {\bibfield  {journal} {\bibinfo  {journal} {Phys. Rev. Lett}\
  }\textbf {\bibinfo {volume} {91}},\ \bibinfo {pages} {108102} (\bibinfo
  {year} {2003})}\BibitemShut {NoStop}%
\bibitem [{\citenamefont {Broedersz}\ \emph {et~al.}(2011)\citenamefont
  {Broedersz}, \citenamefont {Mao}, \citenamefont {Lubensky},\ and\
  \citenamefont {MacKintosh}}]{broedersz2011criticality}%
  \BibitemOpen
  \bibfield  {author} {\bibinfo {author} {\bibfnamefont {C.~P.}\ \bibnamefont
  {Broedersz}}, \bibinfo {author} {\bibfnamefont {X.}~\bibnamefont {Mao}},
  \bibinfo {author} {\bibfnamefont {T.~C.}\ \bibnamefont {Lubensky}}, \ and\
  \bibinfo {author} {\bibfnamefont {F.~C.}\ \bibnamefont {MacKintosh}},\
  }\href@noop {} {\bibfield  {journal} {\bibinfo  {journal} {Nat. Phys.}\
  }\textbf {\bibinfo {volume} {7}},\ \bibinfo {pages} {983} (\bibinfo {year}
  {2011})}\BibitemShut {NoStop}%
\bibitem [{\citenamefont {Sharma}\ \emph {et~al.}(2016)\citenamefont {Sharma},
  \citenamefont {Licup}, \citenamefont {Jansen}, \citenamefont {Rens},
  \citenamefont {Sheinman}, \citenamefont {Koenderink},\ and\ \citenamefont
  {MacKintosh}}]{sharma}%
  \BibitemOpen
  \bibfield  {author} {\bibinfo {author} {\bibfnamefont {A.}~\bibnamefont
  {Sharma}}, \bibinfo {author} {\bibfnamefont {A.~J.}\ \bibnamefont {Licup}},
  \bibinfo {author} {\bibfnamefont {K.~A.}\ \bibnamefont {Jansen}}, \bibinfo
  {author} {\bibfnamefont {R.}~\bibnamefont {Rens}}, \bibinfo {author}
  {\bibfnamefont {M.}~\bibnamefont {Sheinman}}, \bibinfo {author}
  {\bibfnamefont {G.~H.}\ \bibnamefont {Koenderink}}, \ and\ \bibinfo {author}
  {\bibfnamefont {F.~C.}\ \bibnamefont {MacKintosh}},\ }\href@noop {}
  {\bibfield  {journal} {\bibinfo  {journal} {Nat. Phys.}\ }\textbf {\bibinfo
  {volume} {12}},\ \bibinfo {pages} {584} (\bibinfo {year} {2016})}\BibitemShut
  {NoStop}%
\bibitem [{\citenamefont {Dahl}\ \emph {et~al.}(2008)\citenamefont {Dahl},
  \citenamefont {Ribeiro},\ and\ \citenamefont {Lammerding}}]{lammerding}%
  \BibitemOpen
  \bibfield  {author} {\bibinfo {author} {\bibfnamefont {K.~N.}\ \bibnamefont
  {Dahl}}, \bibinfo {author} {\bibfnamefont {A.~J.~S.}\ \bibnamefont
  {Ribeiro}}, \ and\ \bibinfo {author} {\bibfnamefont {J.}~\bibnamefont
  {Lammerding}},\ }\href@noop {} {\bibfield  {journal} {\bibinfo  {journal}
  {Circ. Res.}\ }\textbf {\bibinfo {volume} {102}},\ \bibinfo {pages} {1307}
  (\bibinfo {year} {2008})}\BibitemShut {NoStop}%
\bibitem [{\citenamefont {Wu}\ and\ \citenamefont {et~al.}(2018)}]{wirtz}%
  \BibitemOpen
  \bibfield  {author} {\bibinfo {author} {\bibfnamefont {P.-H.}\ \bibnamefont
  {Wu}}\ and\ \bibinfo {author} {\bibnamefont {et~al.}},\ }\href@noop {}
  {\bibfield  {journal} {\bibinfo  {journal} {Nat. Meth.}\ }\textbf {\bibinfo
  {volume} {15}},\ \bibinfo {pages} {491} (\bibinfo {year} {2018})}\BibitemShut
  {NoStop}%
\bibitem [{\citenamefont {Nakamura}\ \emph {et~al.}(2007)\citenamefont
  {Nakamura}, \citenamefont {Osborn}, \citenamefont {Hartemink}, \citenamefont
  {Hartwig},\ and\ \citenamefont {Stossel}}]{filaminA}%
  \BibitemOpen
  \bibfield  {author} {\bibinfo {author} {\bibfnamefont {F.}~\bibnamefont
  {Nakamura}}, \bibinfo {author} {\bibfnamefont {T.}~\bibnamefont {Osborn}},
  \bibinfo {author} {\bibfnamefont {C.}~\bibnamefont {Hartemink}}, \bibinfo
  {author} {\bibfnamefont {J.}~\bibnamefont {Hartwig}}, \ and\ \bibinfo
  {author} {\bibfnamefont {T.}~\bibnamefont {Stossel}},\ }\href@noop {}
  {\bibfield  {journal} {\bibinfo  {journal} {J. Cell Biol.}\ }\textbf
  {\bibinfo {volume} {179}},\ \bibinfo {pages} {1011} (\bibinfo {year}
  {2007})}\BibitemShut {NoStop}%
\bibitem [{\citenamefont {Liu}\ \emph {et~al.}(2006)\citenamefont {Liu},
  \citenamefont {Jawerth}, \citenamefont {Sparks}, \citenamefont {Falvo},
  \citenamefont {Hantgan}, \citenamefont {Superfine},\ and\ \citenamefont
  {M.}}]{fibrin}%
  \BibitemOpen
  \bibfield  {author} {\bibinfo {author} {\bibfnamefont {W.}~\bibnamefont
  {Liu}}, \bibinfo {author} {\bibfnamefont {L.~M.}\ \bibnamefont {Jawerth}},
  \bibinfo {author} {\bibfnamefont {E.~A.}\ \bibnamefont {Sparks}}, \bibinfo
  {author} {\bibfnamefont {M.~R.}\ \bibnamefont {Falvo}}, \bibinfo {author}
  {\bibfnamefont {R.~R.}\ \bibnamefont {Hantgan}}, \bibinfo {author}
  {\bibfnamefont {S.~T.}\ \bibnamefont {Superfine}, \bibfnamefont {R.~Lord}}, \
  and\ \bibinfo {author} {\bibfnamefont {G.}~\bibnamefont {M.}},\ }\href@noop
  {} {\bibfield  {journal} {\bibinfo  {journal} {Science}\ }\textbf {\bibinfo
  {volume} {313}},\ \bibinfo {pages} {634} (\bibinfo {year}
  {2006})}\BibitemShut {NoStop}%
\bibitem [{\citenamefont {Kim}\ \emph {et~al.}(2014)\citenamefont {Kim},
  \citenamefont {Litvinov}, \citenamefont {Weisel},\ and\ \citenamefont
  {Alber}}]{kim}%
  \BibitemOpen
  \bibfield  {author} {\bibinfo {author} {\bibfnamefont {O.~V.}\ \bibnamefont
  {Kim}}, \bibinfo {author} {\bibfnamefont {R.~I.}\ \bibnamefont {Litvinov}},
  \bibinfo {author} {\bibfnamefont {J.~W.}\ \bibnamefont {Weisel}}, \ and\
  \bibinfo {author} {\bibfnamefont {M.~S.}\ \bibnamefont {Alber}},\ }\href@noop
  {} {\bibfield  {journal} {\bibinfo  {journal} {Biomaterials}\ }\textbf
  {\bibinfo {volume} {35}},\ \bibinfo {pages} {6739} (\bibinfo {year}
  {2014})}\BibitemShut {NoStop}%
\bibitem [{\citenamefont {Fernandez}\ and\ \citenamefont {Ott}(2008)}]{ott}%
  \BibitemOpen
  \bibfield  {author} {\bibinfo {author} {\bibfnamefont {P.}~\bibnamefont
  {Fernandez}}\ and\ \bibinfo {author} {\bibfnamefont {A.}~\bibnamefont
  {Ott}},\ }\href@noop {} {\bibfield  {journal} {\bibinfo  {journal} {Phys.
  Rev. Lett.}\ }\textbf {\bibinfo {volume} {100}},\ \bibinfo {pages} {238102}
  (\bibinfo {year} {2008})}\BibitemShut {NoStop}%
\bibitem [{\citenamefont {Koster}\ \emph {et~al.}(2013)\citenamefont {Koster},
  \citenamefont {Weitz}, \citenamefont {Goldman}, \citenamefont {Aebi},\ and\
  \citenamefont {Herrmann}}]{koster}%
  \BibitemOpen
  \bibfield  {author} {\bibinfo {author} {\bibfnamefont {S.}~\bibnamefont
  {Koster}}, \bibinfo {author} {\bibfnamefont {D.~A.}\ \bibnamefont {Weitz}},
  \bibinfo {author} {\bibfnamefont {R.~D.}\ \bibnamefont {Goldman}}, \bibinfo
  {author} {\bibfnamefont {U.}~\bibnamefont {Aebi}}, \ and\ \bibinfo {author}
  {\bibfnamefont {H.}~\bibnamefont {Herrmann}},\ }\href@noop {} {\bibfield
  {journal} {\bibinfo  {journal} {Curr. Opin. Cell Biol.}\ }\textbf {\bibinfo
  {volume} {32}},\ \bibinfo {pages} {82} (\bibinfo {year} {2013})}\BibitemShut
  {NoStop}%
\bibitem [{\citenamefont {Cartagena-Rivera}\ \emph {et~al.}(2016)\citenamefont
  {Cartagena-Rivera}, \citenamefont {Logue}, \citenamefont {Waterman},\ and\
  \citenamefont {Chadwick}}]{waterman}%
  \BibitemOpen
  \bibfield  {author} {\bibinfo {author} {\bibfnamefont {A.~X.}\ \bibnamefont
  {Cartagena-Rivera}}, \bibinfo {author} {\bibfnamefont {J.~S.}\ \bibnamefont
  {Logue}}, \bibinfo {author} {\bibfnamefont {C.~M.}\ \bibnamefont {Waterman}},
  \ and\ \bibinfo {author} {\bibfnamefont {R.~S.}\ \bibnamefont {Chadwick}},\
  }\href@noop {} {\bibfield  {journal} {\bibinfo  {journal} {Biophys. J.}\
  }\textbf {\bibinfo {volume} {110}},\ \bibinfo {pages} {2528} (\bibinfo {year}
  {2016})}\BibitemShut {NoStop}%
\bibitem [{\citenamefont {Feric}\ and\ \citenamefont
  {Brangwynne}(2013)}]{feric}%
  \BibitemOpen
  \bibfield  {author} {\bibinfo {author} {\bibfnamefont {M.}~\bibnamefont
  {Feric}}\ and\ \bibinfo {author} {\bibfnamefont {C.~P.}\ \bibnamefont
  {Brangwynne}},\ }\href@noop {} {\bibfield  {journal} {\bibinfo  {journal}
  {Nat. Cell Biol.}\ }\textbf {\bibinfo {volume} {15}},\ \bibinfo {pages}
  {1253} (\bibinfo {year} {2013})}\BibitemShut {NoStop}%
\bibitem [{\citenamefont {Patteson}\ \emph
  {et~al.}(2019{\natexlab{b}})\citenamefont {Patteson}, \citenamefont {Pogoda},
  \citenamefont {Byfield}, \citenamefont {Mandal}, \citenamefont
  {Ostrowska-Podhorodecka}, \citenamefont {Charrier}, \citenamefont {Galie},
  \citenamefont {Deptu{\l}a}, \citenamefont {Bucki}, \citenamefont {McCulloch}
  \emph {et~al.}}]{patteson2}%
  \BibitemOpen
  \bibfield  {author} {\bibinfo {author} {\bibfnamefont {A.~E.}\ \bibnamefont
  {Patteson}}, \bibinfo {author} {\bibfnamefont {K.}~\bibnamefont {Pogoda}},
  \bibinfo {author} {\bibfnamefont {F.~J.}\ \bibnamefont {Byfield}}, \bibinfo
  {author} {\bibfnamefont {K.}~\bibnamefont {Mandal}}, \bibinfo {author}
  {\bibfnamefont {Z.}~\bibnamefont {Ostrowska-Podhorodecka}}, \bibinfo {author}
  {\bibfnamefont {E.~E.}\ \bibnamefont {Charrier}}, \bibinfo {author}
  {\bibfnamefont {P.~A.}\ \bibnamefont {Galie}}, \bibinfo {author}
  {\bibfnamefont {P.}~\bibnamefont {Deptu{\l}a}}, \bibinfo {author}
  {\bibfnamefont {R.}~\bibnamefont {Bucki}}, \bibinfo {author} {\bibfnamefont
  {C.~A.}\ \bibnamefont {McCulloch}},  \emph {et~al.},\ }\href@noop {}
  {\bibfield  {journal} {\bibinfo  {journal} {Small}\ }\textbf {\bibinfo
  {volume} {15}},\ \bibinfo {pages} {1903180} (\bibinfo {year}
  {2019}{\natexlab{b}})}\BibitemShut {NoStop}%
\bibitem [{\citenamefont {Licup}\ and\ \citenamefont
  {et~al.}(2015)}]{collagen}%
  \BibitemOpen
  \bibfield  {author} {\bibinfo {author} {\bibfnamefont {A.~J.}\ \bibnamefont
  {Licup}}\ and\ \bibinfo {author} {\bibnamefont {et~al.}},\ }\href@noop {}
  {\bibfield  {journal} {\bibinfo  {journal} {PNAS}\ }\textbf {\bibinfo
  {volume} {112}},\ \bibinfo {pages} {9573} (\bibinfo {year}
  {2015})}\BibitemShut {NoStop}%
\bibitem [{\citenamefont {Ramanujan}(1914)}]{ramanujan}%
  \BibitemOpen
  \bibfield  {author} {\bibinfo {author} {\bibfnamefont {S.}~\bibnamefont
  {Ramanujan}},\ }\href@noop {} {\bibfield  {journal} {\bibinfo  {journal}
  {Quart. J. Math}\ }\textbf {\bibinfo {volume} {45}},\ \bibinfo {pages} {350}
  (\bibinfo {year} {1914})}\BibitemShut {NoStop}%
\bibitem [{\citenamefont {Ferone}\ \emph {et~al.}(2016)\citenamefont {Ferone},
  \citenamefont {Kawohl},\ and\ \citenamefont {Nitsch}}]{ferone2016elastica}%
  \BibitemOpen
  \bibfield  {author} {\bibinfo {author} {\bibfnamefont {V.}~\bibnamefont
  {Ferone}}, \bibinfo {author} {\bibfnamefont {B.}~\bibnamefont {Kawohl}}, \
  and\ \bibinfo {author} {\bibfnamefont {C.}~\bibnamefont {Nitsch}},\
  }\href@noop {} {\bibfield  {journal} {\bibinfo  {journal} {Mathematische
  Annalen}\ }\textbf {\bibinfo {volume} {365}},\ \bibinfo {pages} {987}
  (\bibinfo {year} {2016})}\BibitemShut {NoStop}%
\end{thebibliography}%

\appendix
\section{Cell as a viscous medium surrounded by a cortex}\label{oneloop_appendix}
\subsection{Compressing a 4-gon}\label{comp_4gon}
\begin{figure}[htp]
 \includegraphics[width=0.15\textwidth]{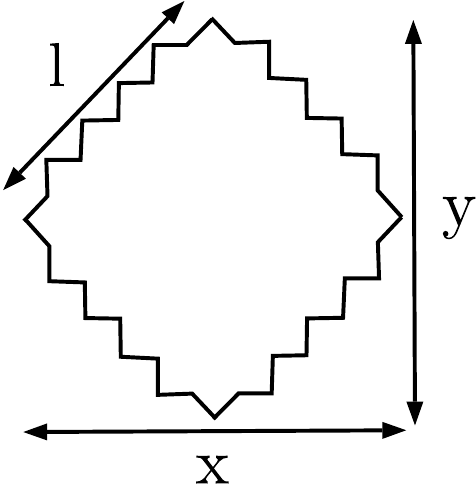}
 \caption{A rhombus 4-gon compressed along the vertical $y-$ axis.}\label{diamond}
\end{figure}

We assume the simplest loop symmetric about the $x-$axis and $y-$axis, a 4-gon that will be vertically and uniaxially compressed.  We choose the arms of the loop to be oblique to the vertical compression rather than to have the arms perfectly parallel to the direction of compression. The latter configuration has the peculiarity that the $x$ degrees of freedom do not couple with the $y$ degrees of freedom when compressed which makes it a non-generic shape. The loop (see Fig. \ref{diamond}) is assumed to be a square at zero strain and a rhombus (all sides equal) at finite strain. 

The central force energy, $H_{o+c,cf}$ is, 
\begin{flalign}\label{2bodyenergy}
  H_{o+c,cf}&=4\;\frac{K_{cf}}{2}\left(l-\frac{1}{\sqrt{2}}\right)^2\nonumber\\
&=2K_{cf}\left\{\sqrt{\left(\frac{x}{2}\right)^2+\left(\frac{y}{2}\right)^2}-\frac{1}{\sqrt{2}}\right\}^2,
 \end{flalign} 
where the rest length of the springs is chosen to be $1/\sqrt{2}$. Since the 4-gon is compressed along the $y-$axis, we define $y=1-\epsilon$ with compressive strain $\gamma=\epsilon/y_0$. The area of the square is $1/2$. Since area is conserved during compression,
$$4\;\frac{1}{2}\left(\frac{x}{2}\right)\left(\frac{y}{2}\right)=1/2,$$
we have,
\begin{equation}\label{xx}
 x=\frac{1}{y}.
\end{equation}
\noindent Substituting Eq. (\ref{xx}) in Eq. (\ref{2bodyenergy}), 
\begin{flalign*}
H_{o+c,cf}&=2K_{cf}\left\{\sqrt{\left(\frac{1}{2y}\right)^2+\left(\frac{y}{2}\right)^2}-\frac{1}{\sqrt{2}}\right\}^2\\
&=K_{cf}\left\{\sqrt{\frac{1}{(1-\epsilon)^2}+(1-\epsilon)^2}-\sqrt{2}\right\},
 \end{flalign*}
where $\epsilon$ is substituted for $y$. The low strain expansion of this energy is,
\begin{equation}\label{diamond-stretch}
 H_{o+c,cf}=2K_{cf}(\epsilon^4+2\epsilon^5+...),
\end{equation}
with $\epsilon\propto \gamma$ and implying compression stiffening.   

With $K_{sf}>0$, a similar calculation can be done with the angular springs to get an expression for the angular spring energy $H_{o+c,sf}$, 
 \begin{equation}\label{ang_expansion}
 H_{o+c,sf}=8\;K_{sf}\left(\epsilon^2+\epsilon^3-\frac{10}{3}\epsilon^4+...\right)
\end{equation}
This latter result gives a quadratic strain term in contrast to the quartic strain term for central-force springs only.  

\subsection{An ellipse in the continuum limit}
When the number of vertices in the loop is large and $\tilde{\kappa}<1$, then numerical minimization yields elliptical shapes with compressive strain. The continuum limit loop is then assumed to be a circle at zero strain and an ellipse at finite strain. For analytical simplicity, a global stretching energy term is used in contrast to a series of individual central force springs in Eq. (\ref{2bodyenergy}). It is seen however that the form of the series expansion of stretching energy is unaffected by this choice (compare Eq. (\ref{diamond-stretch}) and  (\ref{ElStSer})). We have,
\begin{equation}\label{ellipse ham}
 H_{o+c,cf}=\frac{1}{2}K_{cf}(l-l_0)^2.
\end{equation}
An ellipse is defined by two parameters - the semi-major and semi-minor axis which are denoted by $a$ and $b$. The two constraints - the distance between the top and bottom compression walls and the constant area constraint fixes the two parameters of the ellipse, 
\begin{equation}
 \begin{gathered}
  b = 1-\epsilon\\
  \pi ab=\pi r_0^2,
 \end{gathered}
\end{equation}
where $r_o=1$ is the initial state of the loop at zero strain. These constraints reduce the parameters to functions of strain $\epsilon$ as,
\begin{equation}\label{ab}
\begin{gathered}
 a(\epsilon)=\frac{1}{1-\epsilon}\\
 b(\epsilon)=1-\epsilon.
 \end{gathered}
\end{equation}
The circumference $l$ of an ellipse does not have an exact expression and is expressed as the complete elliptic integral of the second kind,
\begin{equation}
 4a\int_0^{\pi/2}\sqrt{1-e^2\;\mbox{sin}^2\theta}\;d\theta,
\end{equation}
where $e=\sqrt{1-b^2/a^2}$ is the eccentricity of the ellipse. Ramanujan's approximation to $l$ is,
\begin{equation}
 l\approx\pi(a+b)\left(1+\frac{3\lambda}{10+\sqrt{4-3\lambda}}\right),
\end{equation}
where $\lambda=(a-b)/(a+b)$. This approximation is good upto $\mathcal{O}(\lambda^{10})$~\cite{ramanujan}. 

Since $a,b,\lambda$ are all expressed as functions of strain $\epsilon,l$, the stretching energy in Eq. (\ref{ellipse ham}) consequently becomes a function of $\epsilon$. The exact expression being dense, a series expansion about zero strain is reproduced here instead,
\begin{equation}\label{ElStSer}
\frac{H_{o+c,cf}}{9\pi^2} \approx K_{cf}\left(\frac{\epsilon^4}{4}+\frac{\epsilon^5}{2}+\frac{23\epsilon^6}{32}+...\right)
\end{equation}
The lowest order term is seen to be quartic just as the stretching energy expression for the discrete loop calculation (Eq.(\ref{diamond-stretch}) ).

Now, we analyze the bending contribution with,   
\begin{equation}\label{ellipse_sf}
 H_{o+c,sf}=\frac{\kappa}{2}\int_0 ^l ds \left|\frac{d \hat{t}}{d s}\right|^2.
\end{equation}
The normal and tangent vector at a point on the ellipse, parameterized by $\theta$ is,
\begin{equation}
 \begin{split}
  \vec{n}&=(a\;\mbox{cos}(\theta),b\;\mbox{sin}(\theta))\\
  \vec{t}&=(-b\;\mbox{sin}(\theta),a\;\mbox{cos}(\theta)).
 \end{split}
\end{equation}
It can be verified that $\vec{n}.\vec{t}=0$. The unit tangent vector $\hat{t}$ is defined as,
\begin{equation}
 \hat{t}=\frac{\vec{t}}{r(\theta)},
\end{equation}
where $r(\theta)$ is,
\begin{equation}
 r(\theta)=(a^2\mbox{cos}^2(\theta)+b^2\mbox{sin}^2(\theta))^{\frac{1}{2}}.
\end{equation}
The contour derivative of the unit tangent vector can be expressed in terms of the parameter $\theta$ as,
\begin{equation}
 \frac{d\hat{t}}{ds}=\frac{1}{r(\theta)}\frac{d\hat{t}}{d\theta}.
\end{equation}
Eq. (\ref{ellipse_sf}) can now be presented as, 
\begin{equation}
H_{o+c,sf} =\frac{\kappa}{2}\int_0^{2\pi}\frac{d\theta}{r(\theta)}\left|\frac{d}{d\theta}\frac{(-b\;\mbox{sin}(\theta),a\;\mbox{cos}(\theta))}{r}\right|^2.
\end{equation}
Having $a,b$ as functions of $\epsilon$, (see Eq. (\ref{ab})), a series expansion of the energy about strain $\epsilon$ can be performed. Subsequent integration over $\theta$ gives,
\begin{equation}
  \frac{H_{o+c,sf}}{\pi}\approx \kappa\left(1+\frac{15}{4}\epsilon^2+\frac{15}{4}\epsilon^3+\frac{405}{64}\epsilon+...\right)
\end{equation}
The form of energy for is again similar to the discrete loop calculation (Eq. (\ref{ang_expansion}). Even though a circle minimizes the bending energy \cite{ferone2016elastica} in Eq. (\ref{ellipse_sf}), it doesn't have zero energy. Thus, we have a constant term here, independent of strain in the energy expansion. 

\subsection{Soft area constraint}\label{soft_area}

We now study the effect of replacing the Lagrange multiplier term in Eq. 1 with a soft area constraint, i.e. $\kappa_A(A-A_0)^2$.  For small enough $K_A$, the area of the semiflexible polymer loop can change and so we ask whether or not compression stiffening will be observed.  For large enough values of $K_A$, we still observe compression stiffening despite changes in area.  See Fig. 9. As discussed in the text, the change in area represents fluid flow from one region of the cell to another. We obtain good agreement with the experimental data with the soft area constraint, suggesting that neither approach, the Lagrange multiplier nor the soft area constraint can yet be ruled out.  Note that $\tilde{\kappa}$ changes modestly from one approach to the other for the experimental comparison.  Finally, the onset of compression stiffening becomes increasingly delayed as $K_A$ goes to zero.  
\begin{figure}[h!]
\includegraphics[width=0.49\textwidth]{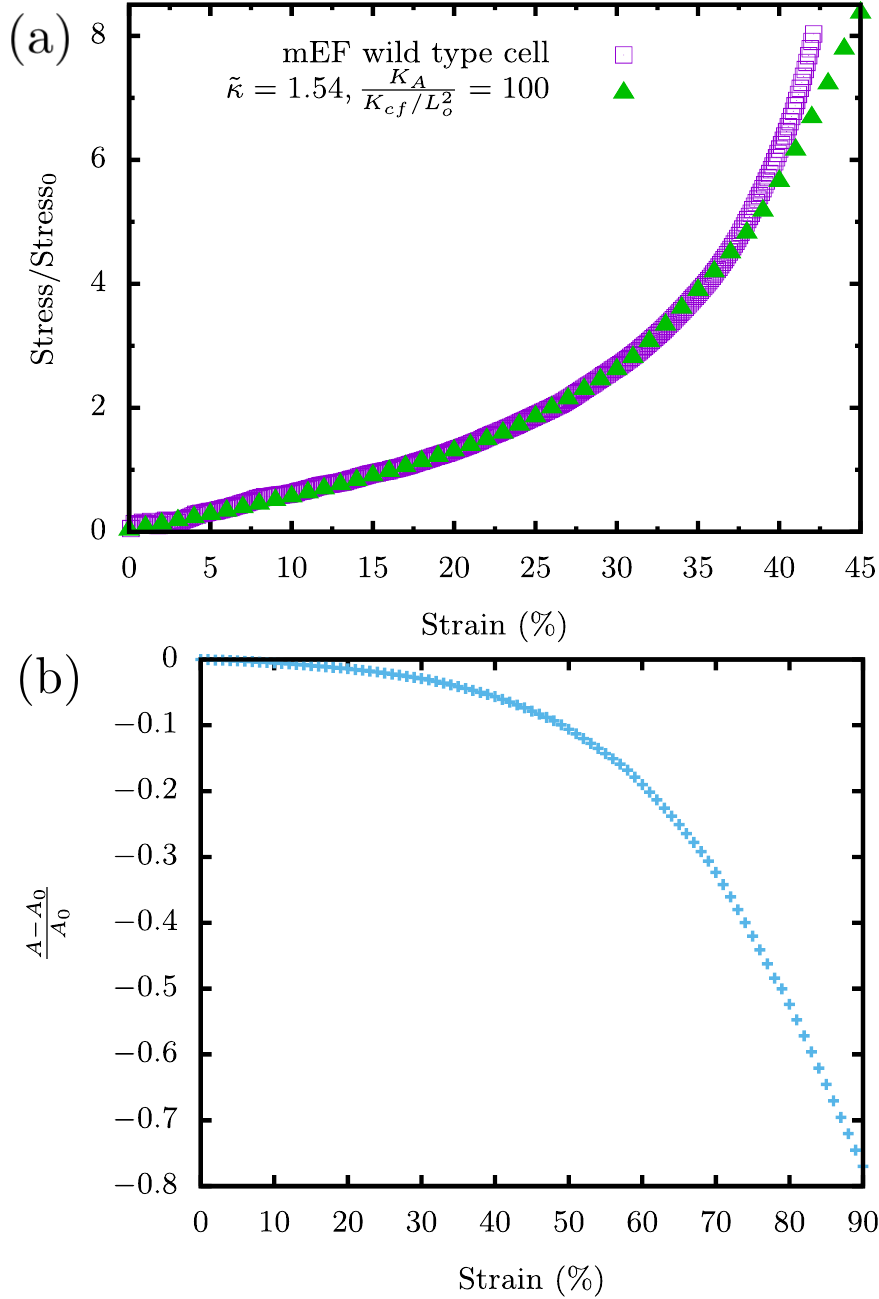} 
\caption{{\it A cell as a viscous interior surrounded by an actomyosin cortex with a soft area constraint.} (a) Plot of the normalized stress versus strain curve from the experiments and the resulting fit. (b) Plot of the corresponding fractional change in area as a function of the compressive strain in the model. }\label{soft area constraint}
\end{figure}

\section{Cell as a collection of organelles within a fiber network}\label{network_appendix}
\subsection{No organelles: Compression softening}\label{comp_soft_appendix}
\begin{figure}[hb]
\includegraphics[width=3 cm]{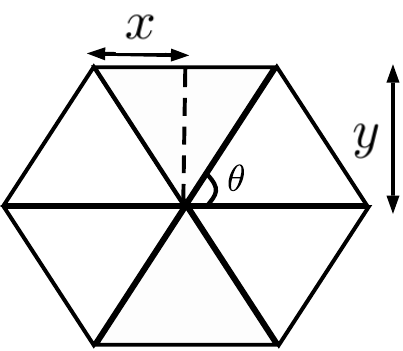} 
 \caption{Collapse of springs induces softening.}\label{collapsesprings}
\end{figure}

\begin{figure*}[ht]
\includegraphics[width = 0.99\textwidth]{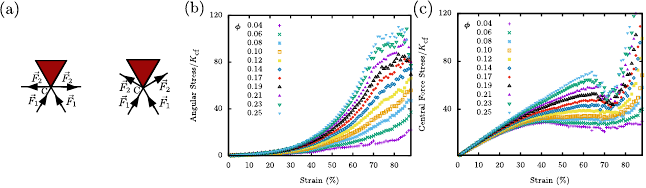}
\caption{\textit{Area conserving loop in a semiflexible polymer network.} (a) Area conserving loop initiates bending (see Appendix \ref{area_initiate_bend}) (b, c) Stress contributed by central force and angular springs respectively, for various packing fractions on a 12x12 lattice with $K_{cf}l_0^2/K_{sf}=1$.}\label{inclusions_figure}
\end{figure*}

\begin{figure}[ht]
\includegraphics[width = 0.49\textwidth]{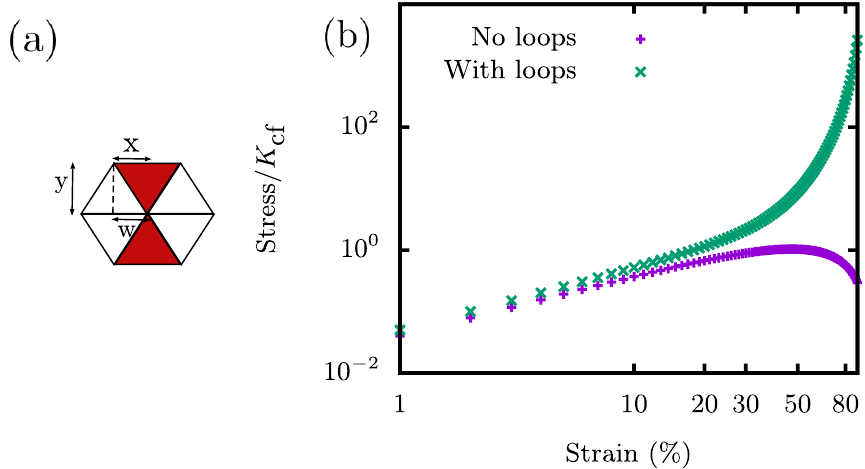}
\caption{\textit{A minimal analytical calculation of an alternate mechanism for compression stiffening which involves a percolation of area-conserving loops and does not require bending.} (a) Schematic  (see Appendix (\ref{area_initiate_bend})).(b) Comparison of analytical calculations with and without loops (see Appendix \ref{comp_4gon}, \ref{area_initiate_bend}).}\label{percolation}
\end{figure}

We first present an approximate calculation for the compression softening mechanism in the absence of organelles (area-conserving loops). The energy of a single central force spring is
\begin{equation}
 E = \frac{K_{cf}}{2}(l-l_0)^2.
\end{equation}
Its differential is
\begin{equation}\label{diff E}
 \Delta E = K_{cf}(l-l_o)\Delta l.
\end{equation}

\begin{figure}[b]
\includegraphics[width = 0.49\textwidth]{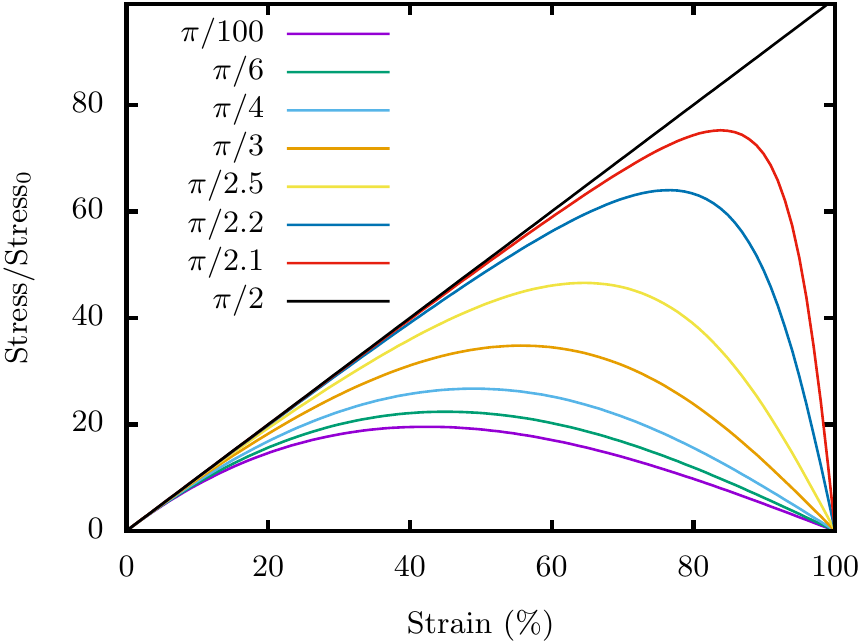}
\caption{\textit{Compression softening is generic to choice of spring orientation in network.}  We study a triangular lattice where the diagonal central force springs make an angle of $\pi/3$ with the transverse axis of compression. Compression softening however is generic non-linear behavior of a central force spring and is seen for all choices of initial orientations of springs with the exception of $\pi/2$ orientation; here the compression being along the axis of the spring, a linear behaviour is observed.  In calculating the curves, we have followed the procedure laid out in the first part of Appendix \ref{comp_soft_appendix}. Stress has been normalized so that the curves overlap at small strain.}\label{comp_soft_angle_dependence}
\end{figure}

\noindent From geometry (see Fig. \ref{collapsesprings}), 
\begin{flalign*}
 l^2&=x^2+y^2\\
 \Rightarrow l\Delta l&\approx y\Delta y\\
 \Rightarrow l\Delta l&\approx l\;\mbox{sin}(\theta)\Delta y.
\end{flalign*}
Assuming $\Delta x\approx 0$, 
\begin{equation}
 \Delta l \approx \mbox{sin}(\theta) \Delta y.
\end{equation}
Substituting the same in Eq. B2, we obtain 
\begin{flalign*}
\Delta E &\approx K_{cf}(l-l_o)\;\mbox{sin}(\theta)\Delta Y\\ 
\frac{\Delta E}{\Delta Y} &\approx K_{cf}(l-l_o)\;\mbox{sin}(\theta).
\end{flalign*}
Since $\Delta y$ is nothing but the strain imposed, using the definition of stress $\sigma$ in Eq. (\ref{stress definition}), 
\begin{equation}
 \sigma \propto \mbox{sin}(\theta).
\end{equation}
As $\theta$ decreases for affine response under compression, stress too decreases. 

\begin{figure*}[ht]
  \includegraphics[width=0.99\textwidth]{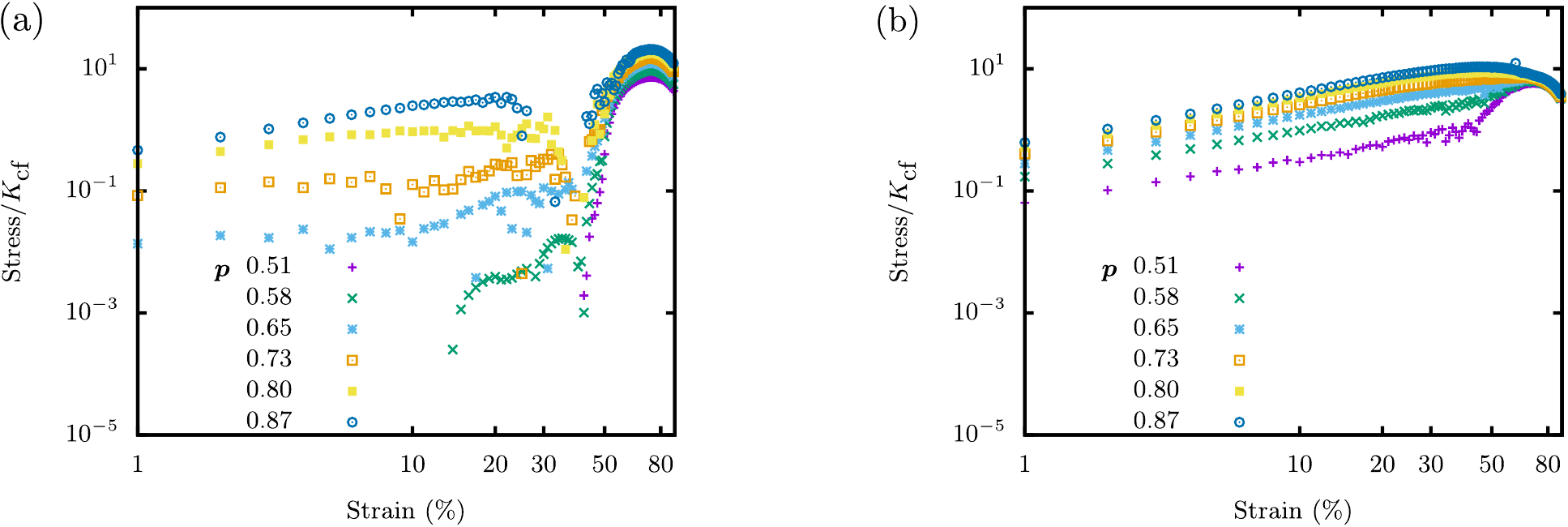}
\caption{\textit{Compression softening observed for occupation probabilities $p<1$.} The curves are averaged over 100 runs on an $8\times8$ lattice obeying the Hamiltonian $H_{o+fn}$ with $K_Al_0^2/K_{cf}=0$. (a) Here, $K_{cf}l_0^2/K_{sf}=0$. (b) Here, $K_{cf}l_0^2/K_{sf}=1$.}
\label{comp softening}
\end{figure*}

\begin{figure}[hb]
\includegraphics[width=0.48\textwidth]{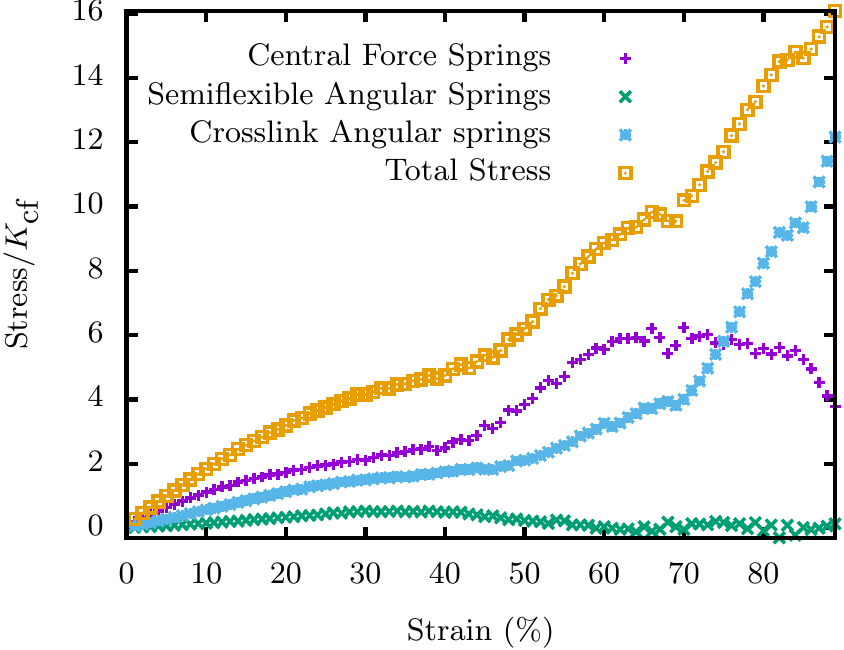}
    \caption{{\it Energetic contributions for angle-constraining crosslink fiber network. } The energy is governed by $H_{fn+axlinks}$ with $p=0.58$, $K_{cf}l_0^2/K_{sf}= 1$, and $K_{sf}/K_{xlink}=10$. The curves are averaged over 100 such initial configurations.}\label{axlink_contributions}
  \end{figure}

We now present a more detailed calculation that makes an exact fit with the numerical results. For an affine deformation, angular springs do not contribute to the elastic energy since straight lines remain straight lines and do not bend. The energy of the system is then the energy of the central force springs in the hexagon (see Fig. \ref{collapsesprings}),
\begin{equation}
 E_o(x,y)\equiv 4\left(\frac{1}{2}\; (2x-1)^2\right)+8\left(\frac{1}{2}\; (\sqrt{x^2+y^2}-1)^2\right).
\end{equation}
The equilibrium lengths of the springs are of unit length. The integer coefficients for the terms are the number of springs that are horizontal and diagonal respectively. For a given compressive strain, the vertical degree of freedom - $y$ is fixed. $E_0$ is minimized over the horizontal degree of freedom - $x$ for every $y$ using \texttt{Mathematica} . This results in the the elastic energy $E_0(y)$ which is now a function of y. The stress is evaluated by taking a derivative with strain to arrive at the plot in Fig. \ref{cellasnetwork_figure}b.

\subsection{Area-conserving loops initiate bending}\label{area_initiate_bend}
Consider the forces acting on vertex $C$ (see Fig. \ref{inclusions_figure}a) in the vertical direction. The summation of the forces must add up to zero to ensure mechanical equilibrium of this vertex. Let us assume that the loop conserves its area by conserving the lengths of each of its side. This implies that the central force springs around the area-conserving loop remain inactive and do not impose any force on vertex $C$. The central force springs directly below the vertex, being compressed, push upward on the vertex. The horizontal springs pull the vertex horizontally as a consequence of Poisson's effect. To balance the upward force on the vertex, the horizontal springs need to bend towards each other. The vertical components of $\vec{F}_2$ would then balance the vertical components of $\vec{F}_1$.

When area-conserving loops are embedded in the network, the network deforms in a non-affine manner. This calculation describes the non-affinity when the said inclusions percolate in the network. The non-affinity in the horizontal degrees of freedom is considered but not the vertical which is an equally important factor to consider. Considering just the energy of the central force springs in the hexagon (see Fig. \ref{percolation}a),
\begin{equation}
\begin{split}
 E_2(x,w,y)&\equiv 2\left(\frac{1}{2}\; (2x-1)^2\right)+2\left(\frac{1}{2}\; (2w-1)^2\right)\\
 &+4\left(\frac{1}{2}\; (\sqrt{x^2+y^2}-1)^2\right)\\
&+4\left(\frac{1}{2}\; (\sqrt{w^2+y^2}-1)^2\right).
\end{split}
\end{equation}
The non-affinity in the horizontal degrees of freedom of the system is captured by assigning two independent variables $x,w$. This energy function has an additional variable calling for an additional constraint to fix its value, which is provided by the area-conserving constraint of the loops, or $\frac{1}{2}\;2x\;y=\frac{1}{2}\times1\times sin(\pi/3)$. The area of the loop at every strain is fixed by the area of the loop at zero strain. With this $E_2$ can be reduced to a function of $x,y$. The rest of the procedure to obtain stress curves is the same as in the no organelle/loop case. Also note that with this geometry the area-conserving loops percolate between the upper and lower plates of the system at the outset, which constrains the deformation of the loops.  This is yet another compression stiffening mechanism that occurs even in the absence of bending and could be very relevant for the reconstituted fibrin network experiments.

\begin{figure}[H]
\includegraphics[width=0.48\textwidth]{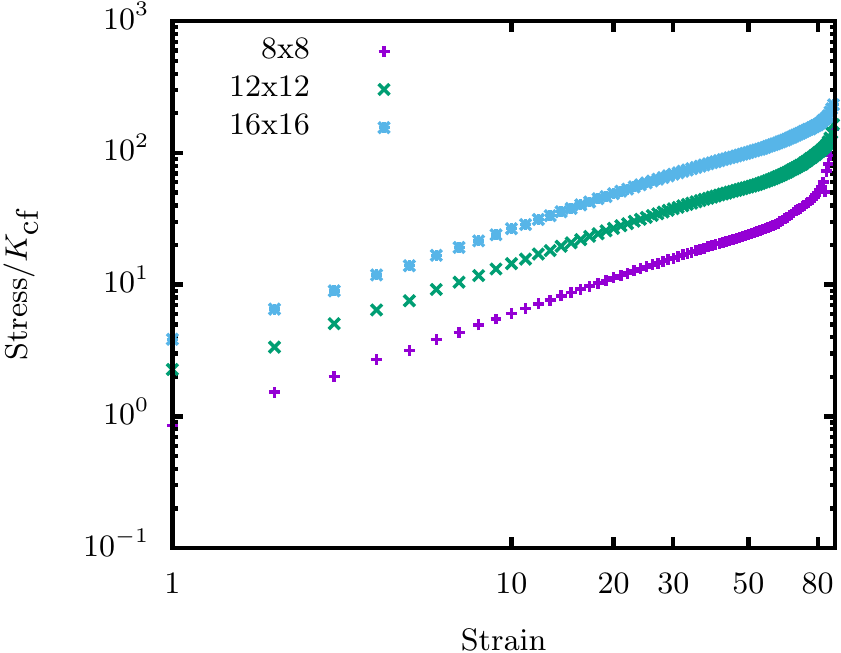}
    \caption{{\it Finite-size effects for fiber network with area-conserving loops. } The energy is governed by $H_{o+fn}$ and the packing fraction of the area-conserving loops are kept constant at $\phi=0.06$ across the lattices. Occupation probability $p=1$ and $K_{cf}l_0^2/K_{sf} = 1$. The position of the loops being random, the curves are averaged over 100 such initial configurations.}\label{finite_size_inclusions}
  \end{figure}    
While the above calculations are representative of the ordered fiber network ($p=1$), we also present some additional numerical results for $p<1$ without and with semiflexibility and in the presence of area-conserving loops.  See Fig. (\ref{comp softening}).  We have also checked that the compressional stiffening persists in both larger and smaller systems and it does with the magnitude of the stress converging as the system size increases and $\gamma_c$ shifting as well with system size. See Fig. (\ref{finite_size_inclusions}).

For the angle-constraining crosslinked fiber network, we present a figure (Fig. \ref{axlink_contributions}) that shows the different energy contributions for each type of spring.  Note that the angular springs along the fibers modeling the semiflexibility do not account for much of the energy even at large compressive strains.

\subsection{Experiment with polyacrylamide gel}

To study the effect of bending in the fiber network on compression stiffening, we study beads embedded in a polyacrylamide (PAA) gel, which is a linear elastic material. The experimental protocol is the following: 8\% acrylamide and 0.3\% bis-acrylamide cross-linker (BioRad, Hercules, CA) was mixed with 10\% ammonium persulfate and TEMED to initiate polymerization, after which it was quickly mixed with pre-swollen G-25 dextran beads and water to produce a network with 2.4\% acrylamide, 0.09\% bis-acrylamide, 0.2\% APS, 0.3\% TEMED, and 40\%, 50\%, or 60\% beads. Then, 1 or 2mm thickness samples were incubated in a non-adhesive container at room temperature for 45 minutes. After full polymerization, samples were cut to size, transferred to the rheometer plates and surrounded by water.

\begin{figure}[ht]
\includegraphics[width=0.48\textwidth]{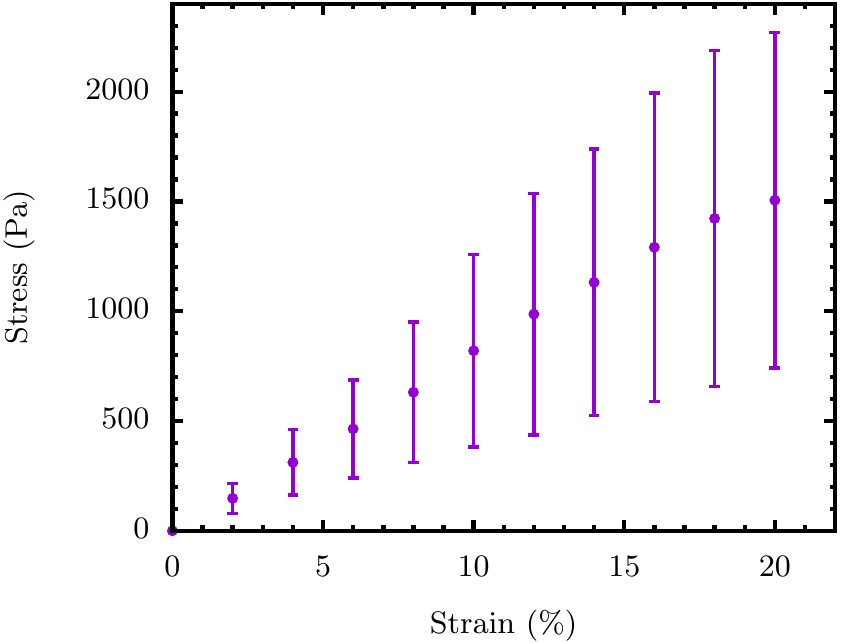}
    \caption{{\it No compression stiffening for PAA gel with beads.}  Plot of the compressive stress versus compressive strain for 60\% dextran beads embedded in a PAA gel.}\label{PAA}
  \end{figure}

 We present data for a 2.4\% PAA gel with 60\% dextran beads and do not find evidence for compression stiffening. See Fig. \ref{PAA}.

\end{document}